\documentclass[aps,prd,preprintnumbers,superscriptaddress,showkeys,showpacs,byrevtex,
fleqn]{revtex4-1}
\usepackage{amsmath,amsfonts,amssymb,amscd,amsxtra,amsthm}
\usepackage{graphicx}
\usepackage{bm}
\usepackage{multirow}
\usepackage{enumerate}

\usepackage[normalem]{ulem} 
\usepackage[dvips]{color} 
\renewcommand\sout{\bgroup \color{red} \ULdepth=-.5ex \ULset}

\begin{document}   
\preprint{KIAS-P12072}
\preprint{INHA-NTG-05/2012}
\title{$K^*\Sigma$ photoproduction off the proton target with
 baryon resonances}        
\author{Sang-Ho Kim}
\email[E-mail: ]{shkim@rcnp.osaka-u.ac.jp}
\affiliation{Research Center for Nuclear Physics (RCNP), Osaka
  567-0047, Japan} 
\affiliation{Department of Physics, Inha University, Incheon 402-751,
  Republic of Korea}  
\author{Seung-il Nam}
\email[E-mail: ]{sinam@kias.re.kr, sinam@pknu.ac.kr}
\affiliation{School of Physics, Korea Institute for Advanced Study
  (KIAS), Seoul 130-722, Republic of Korea} 
\affiliation{Department of Physics, Pukyong National University
  (PKNU), Busan 608-737, Republic of Korea}  
\author{Atsushi Hosaka}
\email[E-mail: ]{hosaka@rcnp.osaka-u.ac.jp}
\affiliation{Research Center for Nuclear Physics (RCNP), Osaka
  567-0047, Japan} 
\author{Hyun-Chul Kim}
\email[E-mail: ]{hchkim@inha.ac.kr}
\affiliation{Department of Physics, Inha University, Incheon 402-751, 
  Republic of Korea}
\affiliation{School of Physics, Korea Institute for Advanced Study
  (KIAS), Seoul 130-722, Republic of Korea} 
\date{\today}
\begin{abstract}
We investigate the photoproduction of $K^{*0}\Sigma^+$ and
$K^{*+}\Sigma^0$ off the proton target, employing the effective
Lagrangian approach at the tree-level Born approximation. In addition
to the $(s,t,u)$-channel Born diagrams, we take into account
various baryon-resonance contributions such as  
$D_{13}(2080)$, $S_{11}(2090)$, $G_{17}(2190)$, $D_{15}(2200)$, 
$S_{31}(2150)$, $G_{37}(2200)$, $F_{37}(2390)$, 
and $\Sigma^*(1385,3/2^+)$ in a fully covariant manner. 
We present the numerical results for the energy
and angular dependences for the cross sections in comparison to
available experimental data. The single-polarization observables,
i.e. the photon-beam ($\Sigma_\gamma$), recoil ($P_y$) and target
($T_y$) baryon polarization asymmetries are computed as well for 
future experiments. We observe from the numerical results that the
resonance contributions play a minor role in producing the strength of 
the cross sections, being different from the $K^*\Lambda$
photoproduction. In contrast, it turns out that the
$\Delta(1232)$-pole contribution and 
$K$ exchange in the $t$-channel dominate the scattering process.   
On the other hand, the higher resonances influence the polarization
observables such as the recoil and target asymmetries. 
\end{abstract} 
\maketitle
\section{Introduction}
Strangeness production via various scattering processes has been one
of the most important issues in hadronic and nuclear physics 
for decades. From them, we can understand the microscopic mechanism of
the productions beyond the light-flavor sectors and extend our
knowledge into multistrangeness states. In this sense,
photoproduction of strange hadrons off the nucleon target is a
very useful  tool and has been widely studied
experimentally as well as theoretically. For example, experiments for
the photoproduction of $\gamma N\to K\Lambda$ and $K\Sigma$ were
reported in Refs.~\cite{Bradford:2005pt, Tsukada:2007jy,
  Watanabe:2006bs}. Related theoretical studies were also performed in
Refs.~\cite{Janssen:2001wk,Mart:2011ez,Yu:2011fv,DeCruz:2010ry}. In
particular, Ref.~\cite{Janssen:2001wk} emphasized the baryon-resonance 
contributions, which play important roles in reproducing the
experimental data. The effects of the electromagnetic form
factor~\cite{Mart:2011ez} were also investigated for the photo- and
electroproduction of the kaon, the Ward-Takahashi (WT) identity being
explained. It was also pointed out that the tensor-meson
exchange in the $t$ channel provides a significant contribution to 
kaon photoproduction~\cite{Yu:2011fv}. An unbiased model selection,
based on Bayesian inference, was introduced for extracting physical
information from kaon
photoproduction~\cite{DeCruz:2010ry}. References~\cite{Corthals:2005ce,
  Corthals:2006nz} examined the $t$-channel Regge trajectories to
enhance the model applicabilities to actual problems. 

Photoproduction of the vector strange meson $(K^*)$ provides even
richer physics in comparison with the $KY$ channel. For instance,
since it is a vector meson with quantum number $I(J^P)=1/2(1^-)$, the
exchange of the strange scalar meson $\kappa$ is allowed in the
$t$ channel, which is absent in the $KY$ channel, in addition to 
$(K,K^*)$ exchanges. Moreover, the polarization of the $K^*$ meson in
the final state can be taken as an important  subject to be
investigated together with other polarization observables in terms of
the spin-density matrices. Experimentally, this production channel has 
been investigated for $\gamma N\to K^*\Lambda(1116)$ by the CLAS 
Collaboration at the Thomas Jefferson National Accelerator Facility
(Jefferson Lab)~\cite{Guo:2006kt,Hicks:2010pg}, 
and $\gamma N\to K^*\Sigma(1193)$ by the CBELSA/TAPS Collaboration at
the Electron Stretcher and Accelerator (ELSA)~\cite{Nanova:2008kr}, 
by the CLAS Collaboration~\cite{Hleiqawi:2005sz,Hleiqawi:2007ad},
and by the LEPS Collaboration at Super Photon Ring-8 GeV
(SPring-8)~\cite{Hwang2012}. These two processes have been extensively
studied theoretically within the effective Lagrangian 
approaches~\cite{Oh:2006hm,Ozaki:2009wp,Oh:2006in,Kim:2011rm}, as well 
as in the chiral quark model~\cite{Zhao:2001jw}. As mentioned above,
it was argued that the $\kappa$-exchange should play an important role
in the production mechanism of $\gamma p\to
K^*\Sigma$~\cite{Oh:2006in}. Interestingly enough, the recent 
LEPS experiment reported the experimental data that supported the
importance of the scalar-meson exchange
indeed~\cite{Hwang2012}. Moreover, employing the same theoretical 
framework, Ref.~\cite{Kim:2011rm} showed that
there were some contributions from nucleon resonances
to reproduce the experimental data of $\gamma p\to K^{*+}\Lambda$.
 
Considering all these successful and meaningful theoretical results
accumulated so far within the effective Lagrangian method with the 
resonance contributions taken into account, we want to explore
carefully the reaction processes $\gamma p\to K^{*0}\Sigma^+$  and
$\gamma p\to K^{*+}\Sigma^0$ in the present work. Although the
$K^*\Sigma$ photoproduction was already studied theoretically within
a similar framework in Ref.~\cite{Oh:2006in}, we will include
various baryon-resonance contributions which were proven to be
essential in the $K^*\Lambda$ channel~\cite{Kim:2011rm}. Thus, we
introduce the baryon resonances as follows:  
$D_{13}(2080)$, $S_{11}(2090)$, $G_{17}(2190)$, $D_{15}(2200)$,
$S_{31}(2150)$, $G_{37}(2200)$, and $F_{37}(2390)$ in the 
$s$ channel and 
$\Sigma^*(1385)$ in the $u$ channel, in addition to the $s$ channel
with $N(940)$- and $\Delta(1232)$-pole contributions ; the $t$ channel
with $\kappa$-, $K$-, and $K^*$-exchange contributions ; and the 
$u$ channel with $\Lambda(1116)$- and $\Sigma(1193)$-pole
contributions. These resonance contributions have not been taken into
account in the previous theoretical work~\cite{Oh:2006in} and will be
treated in a fully relativistic manner in the present work, as done
for the $\gamma p\to K^{*+}\Lambda$~\cite{Kim:2011rm}. 

The coupling strengths for strong and electromagnetic (EM) vertices are
computed by using experimental and theoretical
information~\cite{Nakamura:2010zzi, Oh:2008abc, Stoks:1999bz,
Capstick:1992uc, Capstick:1998uh}. 
In order to preserve the WT identity, we employ the gauge-invariant
form factor prescription given in Refs.~\cite{Haberzettl:1998eq,
  Davidson:2001rk, Haberzettl:2006bn}. The cutoff parameters for the
form factors are determined in such a way that the experimental data
are reproduced. With these parameters fixed, we compute the total  
($\sigma$) and differential cross sections $(d\sigma/d\Omega)$ for the
$\gamma p \to K^* \Sigma$ processes. In addition, the single-polarization
observables such as those for the photon-beam $(\Sigma)$,
target $(T_y)$, recoil baryon $(P_y)$, are presented as useful
theoretical guides for available and future experiments. Based on the
present results, we observe that the resonance contributions play a
minor role in producing the strength of the cross sections, being
different from the $K^*\Lambda$ photoproduction. On the other hand, 
it turns out that the $\Delta(1232)$-pole diagram and 
$K$ exchange in the $t$ channel are dominant in explaining the 
production mechanism of $\gamma p \to K^*\Sigma$.  

The present work is organized as follows: In Sec. II, we explain
the general formalism of the effective Lagrangian methods and 
show how to fix various model parameters such as the coupling
constants and the cutoff masses.  The numerical results are presented
and discussed in Sec. III. The last section is devoted to 
the summary, conclusion, and future perspectives.  
\section{Formalism}
\begin{figure}[ht]
\includegraphics[width=12cm]{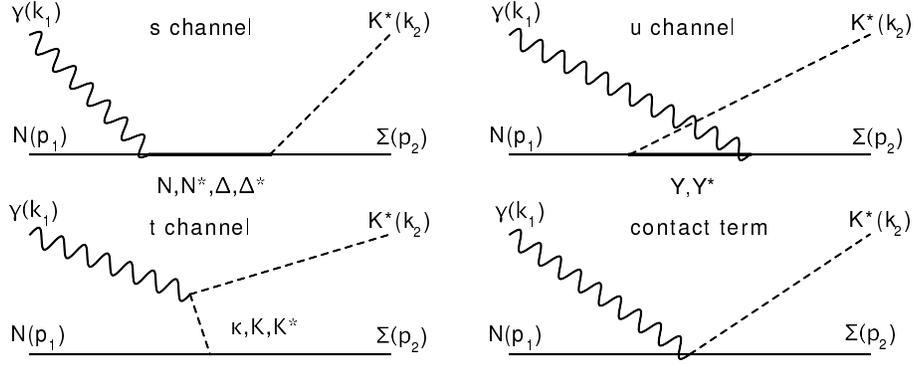}
\caption{Relevant Feynman diagrams for the $\gamma N\to K^*\Sigma$
  reactions. $N$, $N^*$, $\Delta$, $\Delta^*$, $Y$, and $Y^*$
  denote the nucleon, nucleon resonances, delta, delta resonances,
  hyperons, and hyperon resonances, respectively, whereas $\kappa$,
  $K$, and $K^*$ stand for the strange scalar, pseudoscalar, and
  vector mesons, respectively. The four momenta for the initial and
  final states are also defined, as shown in the diagrams.}        
\label{FIG1}
\end{figure}
We start with the effective Lagrangian method at the
tree-level Born approximation. The relevant and generic Feynman
diagrams for the reaction processes $\gamma p\to K^{*0}\Sigma^+$ and
$\gamma p\to K^{*+}\Sigma^0$ are shown in 
Fig.~\ref{FIG1}, which include $N$, $\Delta$, $N^*$, and $\Delta^*$ poles
in the $s$ channel, the $K^*$, $K$, and $\kappa$ meson exchanges in
the $t$ channel, and $\Lambda$, $\Sigma$, and $\Sigma^*(1385,3/2^+)$
hyperons in the $u$ channel. The contact-term contribution is
necessary for satisfying the WT identity. For convenience, we
assign these two production processes as the $K^{*0} \Sigma^+$ and $
K^{*+} \Sigma^0$ channels, respectively, from now on. Note
that, however, we do not have the $K^*$ exchange for the
$K^{*0}\Sigma^+$ channel due to their electrically neutral vertex of
$\gamma K^*\bar{K}^*$ as far as we ignore the magnetic and quadratic
moments of $K^*$ as in the present work. Consequently, the contact
term is also absent for the $K^{*0}\Sigma^+$ channel.  

The effective Lagrangians for the Born contributions are essentially
the same as those used in Refs.~\cite{Oh:2006hm,Kim:2011rm}. As for
the photon-meson-meson interactions, we define them as follows: 
\begin{eqnarray}
\label{eq:LAGEM}
\mathcal L_{\gamma K^*K^*} &=&
-ie_{K^*}A^\mu (K^{*-\nu}K_{\mu\nu}^{*+} -
K_{\mu\nu}^{*-}K^{*+\nu}), 
\cr
\mathcal L_{\gamma K^*K} &=&
g_{\gamma KK^*} \varepsilon^{\mu\nu\alpha\beta} \left( \partial_\mu
  A_\nu \right)  
\left(\partial_\alpha K_\beta^* \right) \bar K + \mathrm{h.c.},
\cr
\mathcal{L}_{\gamma K^* \kappa} &=&
g_{\gamma K^* \kappa} F^{\mu\nu} \bar\kappa K^*_{\mu\nu} + \mathrm{h.c.}, 
\label{eq1}
\end{eqnarray}
where $A_\mu$, $K^*_\mu$, $K$, and $\kappa$ denote the photon, the
$K^*(892,1^-)$, $K(495,0^-)$, and $\kappa(800,0^+)$,
respectively~\cite{Nakamura:2010zzi}. The field-strength tensors for
the photon and the massive vector meson are defined as $F_{\mu\nu}
= \partial_\mu A_\nu - \partial_\nu A_\mu$ and
$K_{\mu\nu}^*=\partial_\mu K^*_\nu-\partial_\nu K^*_\mu$,
respectively. The values for the coupling constants 
$g_{\gamma K^*K}$ are determined from the 
experimental data~\cite{Nakamura:2010zzi}, which lead to  
\begin{equation}
\label{eq:GAMMAKSK}
g^\mathrm{charged}_{\gamma K^*K}=0.254\,\mathrm{GeV}^{-1},\,\,\,\,
g^\mathrm{neutral}_{\gamma K^*K}=-0.388\,\mathrm{GeV}^{-1},
\end{equation}
whereas we use the vector-meson dominance model to determine the
values of $g_{\gamma K^* \kappa}$~\cite{Black:2002ek}:  
\begin{equation}
\label{eq:GAMMAKMKAPPA}
g^\mathrm{charged}_{\gamma
  K^*\kappa}=-0.119\,e\,\mathrm{GeV}^{-1},\,\,\,\,  
g^\mathrm{neutral}_{\gamma K^*\kappa}= -2 g^\mathrm{charged}_{\gamma
  K^*\kappa}. 
\end{equation}
Here $e$ denotes the unit electric charge
$e=\sqrt{4\pi\alpha_{\mathrm{E}}}$ with the fine-structure constant 
$\alpha_{\mathrm{EM}}=1/137.04$.
As for the $\kappa$ meson's parameters, we use $M_\kappa$ = 800
MeV for the mass, and $\Gamma$ = 550 MeV for the decay width value.    

The Lagrangians for the photon-baryon-baryon interactions are
written by  
\begin{eqnarray}
\mathcal L_{\gamma NN}&=&
-\overline{N} \left[ e_N\rlap{\,/}{A} - \frac{e\kappa_N }{2M_N} 
\sigma_{\mu\nu} \partial^\nu A^\mu \right] N,
\cr
\mathcal L_{\gamma N\Delta}&=&
e \overline{\Delta}_\mu \left[ \frac{ig_1}{2M_N} \gamma_\nu \gamma_5  
+\frac{g_2}{(2M_N)^2} \gamma_5 \partial_\nu \right]  
N F^{\mu\nu} + \mathrm{h.c.},
\cr
\mathcal L_{\gamma\Sigma\Sigma}&=&
-\overline{\Sigma} \left[ e_\Sigma\rlap{\,/}{A} - \frac{e\kappa_\Sigma
  }{2M_N}  
\sigma_{\mu\nu} \partial^\nu A^\mu \right] \Sigma,
\label{eq:GBB}
\end{eqnarray}
where $N$, $\Sigma$, and $\Delta$ stand for the nucleon,
$\Sigma(1193,1/2^+)$, and $\Delta(1232,3/2^+)$, respectively, and
$M_N$ denotes the mass of the nucleon. Here $\kappa_B $ represents 
the anomalous magnetic moment of the baryon $B$. The corresponding PDG
values~\cite{Nakamura:2010zzi} are given as
\begin{equation}
\label{eq:ANO}
\kappa_n=-1.91,\,\,\,\,
\kappa_p=+1.79,\,\,\,\,
\kappa_{\Sigma^-}=-0.16,\,\,\,\,
\kappa_{\Sigma^0}=+0.65,\,\,\,\,
\kappa_{\Sigma^+}=+1.46.
\end{equation}
The $\Delta$ field with spin-$3/2$ is described by the
Rarita-Schwinger formalism~\cite{Rarita:1941mf,Read:1973ye}. We
choose the electric and magnetic couplings as $g_1=4.13$ and
$g_2=4.74$ using the experimental data for the helicity amplitudes
~\cite{Nakamura:2010zzi,Oh:2008abc}.   

We define the effective Lagrangians for the meson-baryon-baryon Yukawa  
interactions as follows: 
\begin{eqnarray}
\label{eq:MBB}
\mathcal L_{K^* N\Sigma} &=&
-g_{K^* N \Sigma} \left[ \overline K^{*\mu} \overline \Sigma \gamma_\mu - 
\frac{\kappa_{K^* N\Sigma} }{2M_N} \partial^\nu 
\overline K^{*\mu} \overline \Sigma \sigma_{\mu\nu} 
\right] N + \mbox{h.c.},  
\cr
\mathcal L_{KN\Sigma} &=&
-ig_{KN\Sigma} \overline K\, \overline\Sigma \gamma_5 N + \mathrm{h.c.},       
\cr 
\mathcal{L}_{\kappa N\Sigma} &=&
-g_{\kappa N\Sigma} \overline\kappa \overline\Sigma N + \mathrm{h.c.},     
\cr
\mathcal L_{K^*\Delta\Sigma} &=& 
-\frac{if_{K^*\Delta\Sigma}}{2M_{K^*}}  
\overline{\Delta}^\mu \gamma^\nu \gamma_5 \Sigma K^*_{\mu\nu}
+\mathrm{h.c.},
\end{eqnarray}
where $\Sigma=\bm{\tau}\cdot\bm{\Sigma}$ in which $\bm{\tau}$
indicate the Pauli matrices. The isospin structures of the $\Delta$
vertices in Eqs.~(\ref{eq:GBB}) and ~(\ref{eq:MBB}) are given as
follows, respectively: 
\begin{eqnarray}
\label{eq:ISOSPIN}
\overline\Delta I^0 N,\,\,\, \overline\Delta
\bm{I}\cdot\bm{\Sigma}K^* ,
\end{eqnarray} 
where $\bm I$ stands for the isospin transition ($3/2\to 1/2$) matrices  
\begin{equation}
\label{eq:ISO}
I^- =
\frac{1}{\sqrt{6}}\left(\begin{array}{cc}
0&0\\
0&0\\
\sqrt{2}&0\\
0&\sqrt{6}
\end{array} \right),\,\,\,\,
I^0 =
\frac{1}{\sqrt{6}}\left(\begin{array}{cc}
0&0\\
2&0\\
0&2\\
0&0
\end{array} \right),\,\,\,\,
I^+ =
\frac{1}{\sqrt{6}}\left(\begin{array}{cc}
\sqrt{6}&0\\
0&\sqrt{2}\\
0&0\\
0&0
\end{array} \right).
\end{equation}
The strong coupling constants for the meson and octet baryons can be
estimated by the Nijmegen soft-core model
(NSC97a)~\cite{Stoks:1999bz}, and the corresponding values are
presented by 
\begin{equation}
\label{eq:NSC97a}
g_{K^*N\Sigma}=-2.46,\,\,\,\,
\kappa_{K^*N\Sigma}=-0.47,\,\,\,\,
g_{\kappa N\Sigma}=-5.32,
\end{equation}
whereas we estimate the value of $f_{K^*\Delta\Sigma}$ using the 
quark-model prediction and SU(3) flavor symmetry relation:
\begin{equation}
f_{K^*\Delta\Sigma}=-\frac{2M_{K^*}}{M_\rho}f_{\rho N\Delta}=-12.8 ,
\end{equation}
with $f_{\rho N\Delta}=5.5$~\cite{Machleidt:1987hj}.
The value of $g_{KN\Sigma}$ is also obtained by using a similar
relation,  which gives $g_{KN\Sigma}$ = 3.58.

Now, we are in a position to consider the resonance
contributions. First, we write the EM and strong effective Lagrangians  
with the hyperon resonance $\Sigma^*$: 
\begin{eqnarray}
\label{eq:SIGMASTAR}
\mathcal L_{\gamma\Sigma\Sigma^*}&=&
e \overline{\Sigma}_\mu^* \left[
  \frac{ig^V_{\gamma\Sigma\Sigma^*}}{2M_N} \gamma_\nu \gamma_5 
+\frac{g^T_{\gamma\Sigma\Sigma^*}}{(2M_N)^2} \gamma_5 \partial_\nu
\right]   
\Sigma F^{\mu\nu} + \mathrm{h.c.},
\cr
\mathcal L_{K^* N\Sigma^*} &=& 
i\frac{f^{(1)}_{K^* N\Sigma^*}}{2M_{K^*}} \overline N \gamma^\nu \gamma_5   
\Sigma^{*\mu} K^*_{\mu\nu} + 
\frac{f^{(2)}_{K^* N\Sigma^*}}{(2M_{K^*})^2} \partial^\nu \overline N 
\gamma_5 \Sigma^{*\mu} K^*_{\mu\nu} -
\frac{f^{(3)}_{K^* N\Sigma^*}}{(2M_{K^*})^2} \overline N \gamma_5  
\Sigma^{*\mu} \partial^\nu K^*_{\mu\nu}  
+\mathrm{h.c.}.
\end{eqnarray}
In order to determine $g^{V,T}_{\gamma\Sigma\Sigma^*}$, we need to
know the experimental data for the $\Sigma^*\to \Sigma \gamma$
radiative decay. However, only the upper limits of the hyperon decay
rates are known~\cite{Molchanov:2004iq}. Moreover, $\Sigma^{*-}\to
\Sigma^-\gamma$ is known to be $U$-spin forbidden, which means its
decay rate vanishes in the exact SU(3) symmetry. On the other hand,
these decay rates were predicted within several different theoretical
frameworks~\cite{Hackman:1977am,Darewych:1983, Schat:1995mt,
  Abada:1995db, Wagner:1998bu, Kim:2005gz}. Since
Ref.~\cite{Wagner:1998bu} has computed the hyperon radiative decay rates
as well as the $E2/M1$ ratio, we use the results of
Ref.~~\cite{Wagner:1998bu}, so that we are able to extract
$g^{V,T}_{\gamma\Sigma\Sigma^*}$ as follows: 
\begin{eqnarray}
\label{eq:}
&&g^{V+}_{\gamma\Sigma\Sigma^*}=+2.66,\,\,\,\,g^{T+}_{\gamma\Sigma\Sigma^*}=+0.74, \cr
&&g^{V0}_{\gamma\Sigma\Sigma^*}=+1.10,\,\,\,\,g^{T0}_{\gamma\Sigma\Sigma^*}=+0.55, \cr
&&g^{V-}_{\gamma\Sigma\Sigma^*}=+0.49,\,\,\,\,g^{T-}_{\gamma\Sigma\Sigma^*}=-0.39.
\end{eqnarray}
The coupling constant $f^{(1)}_{K^* N\Sigma^*}$ can be determined to
be $-5.21$ by flavor SU(3) symmetry. Because of the lack of
experimental and theoretical information on $f^{(2,3)}_{K^*
  N\Sigma^*}$, we do not consider them for brevity in the present
work.   

In addition to the hyperon resonances, we now include the $s$-channel 
resonance contributions. Here, we consider the 
$D_{13}(2080)$, $S_{11}(2090)$, $G_{17}(2190)$, and $D_{15}(2200)$ for 
the nucleon and 
$S_{31}(2150)$, $G_{37}(2200)$, and $F_{37}(2390)$ for 
the delta 
resonances, which are located near the threshold of $K^*\Sigma$
photoproduction. The relevant EM Lagrangians for those
baryon resonances can be written as  
\begin{eqnarray}
\mathcal{L}_{\gamma  NR_{1/2^\pm}}  &=& 
\frac{eh_1}{2M_N} \overline{N} \Gamma^{(\mp)}\sigma_{\mu\nu}
\partial^\nu A^\mu R + \mathrm{h.c.},                       
\cr
\mathcal{L}_{\gamma  NR_{3/2^\pm}} &=& 
-ie \left[ \frac{h_{1}}{2M_N} \overline{N} \Gamma_\nu^{(\pm)} - 
           \frac{ih_{2}}{(2M_N)^2} \partial_\nu \overline{N}
           \Gamma^{(\pm)} \right] 
F^{\mu\nu} R_\mu  + \mathrm{h.c.},                       
\cr
\mathcal{L}_{\gamma  NR_{5/2^\pm}}  &=& 
e \left[ \frac{h_{1}}{(2M_N)^2} \overline{N} \Gamma_\nu^{(\mp)} -
         \frac{ih_{2}}{(2M_N)^3} \partial_\nu \overline{N}
         \Gamma^{(\mp)}    \right] 
\partial^\alpha F^{\mu\nu} R_{\mu\alpha} + \mathrm{h.c.},
\cr
\mathcal{L}_{\gamma  NR_{7/2^\pm}}&=& 
ie \left[ \frac{h_{1}}{(2M_N)^3} \overline{N} \Gamma_\nu^{(\pm)} -
          \frac{ih_{2}}{(2M_N)^4} \partial_\nu \overline{N}
          \Gamma^{(\pm)} \right] 
\partial^\alpha \partial^\beta F^{\mu\nu} R_{\mu\alpha\beta} +
\mathrm{h.c.},   
\label{eq:LAGEMRE}
\end{eqnarray}
where $R$ stands for the field corresponding to the nucleon and
delta resonances $R=(N^*,\Delta^*)$ with spin and parity given. 
$\Gamma^{(\pm)}$ and $\Gamma_\nu^{(\pm)}$ in Eq.(\ref{eq:LAGEMRE}) are 
defined as
\begin{equation}
\label{eq:GAMMASPEM}
\Gamma^{(\pm)} = \left(
\begin{array}{c} 
\gamma_5 \\ \mathbf{1}
\end{array} \right),
\qquad
\Gamma_\mu^{(\pm)} = \left(
\begin{array}{c}
\gamma_\mu \gamma_5 \\ \gamma_\mu 
\end{array} \right).
\end{equation}
The coupling constants are determined by using the experimental data
for the helicity amplitudes~\cite{Nakamura:2010zzi,Oh:2008abc} and the
quark-model predictions of
Ref.~\cite{Capstick:1992uc,Oh:2008abc}. Those for the strong 
interactions are given as   
\begin{eqnarray}
\label{eq:STRONG}
\mathcal{L}_{ K^* \Sigma R_{1/2^\pm}}
&=&-\frac{1}{2M_N} \overline R \left[
g_1 \biggl( \pm \frac{\Gamma_\mu^{(\mp)} \Sigma \partial^2}{M_R \mp M_N}
-i \Gamma^{(\mp)} \partial_\mu \biggr)
-g_2 \Gamma^{(\mp)} \sigma_{\mu\nu} \Sigma \partial^\nu 
\right] K^{*\mu} + \mathrm{h.c.},
\cr
\mathcal{L}_{ K^* \Sigma R_{3/2^\pm}}                                    
&=& i\overline R_\mu \left[ 
\frac{g_1}{2M_N} \Sigma \Gamma_\nu^{(\pm)} \mp 
\frac{ig_2}{(2M_N)^2} \partial_\nu \Sigma \Gamma^{(\pm)} \pm 
\frac{ig_3}{(2M_N)^2} \Sigma \Gamma^{(\pm)} \partial_\nu 
\right] K^{*\mu\nu} + \mathrm{h.c.},                                                             
\cr
\mathcal{L}_{ K^* \Sigma R_{5/2^\pm}}                                     
&=& \overline R_{\mu\alpha} \left[ 
\frac{g_1}{(2M_N)^2} \Sigma \Gamma_\nu^{(\mp)} \pm 
\frac{ig_2}{(2M_N)^3} \partial_\nu \Sigma \Gamma^{(\mp)} \mp 
\frac{ig_3}{(2M_N)^3} \Sigma \Gamma^{(\mp)} \partial_\nu 
\right] \partial^\alpha K^{*\mu\nu} + \mathrm{h.c.},                                        
\cr
\mathcal{L}_{ K^* \Sigma R_{7/2^\pm}} 
&=& -i\overline R_{\mu\alpha\beta} \left[ 
\frac{g_1}{(2M_N)^3} \Sigma \Gamma_\nu^{(\pm)} \mp 
\frac{ig_2}{(2M_N)^4} \partial_\nu \Sigma \Gamma^{(\pm)} \pm 
\frac{ig_3}{(2M_N)^4} \Sigma \Gamma^{(\pm)} \partial_\nu 
\right] \partial^\alpha \partial^\beta K^{*\mu\nu} + \mathrm{h.c.}, 
\end{eqnarray}
where $M_R$ is the corresponding resonance mass. The strong coupling
constants in Eq.~(\ref{eq:STRONG}) can be determined from the
theoretical estimations for the partial-wave decay 
amplitudes~\cite{Capstick:1998uh}:
\begin{equation}
\label{eq:GGG}
\Gamma_{R\to K^*\Sigma}=\sum_{s,l}|G(s,l)|^2,
\end{equation}
where $\Gamma_{R\to K^*\Sigma}$ is the decay width of $R \to K^*
\Sigma$. The values for the partial-wave coupling strengths $G(s,l)$
can be found in Ref.~~\cite{Capstick:1998uh}. Since the purpose of the
present work is to investigate the role of resonances near the
threshold, it is enough to take into account the contributions of 
the lower partial waves. Hence, we consider only the $g_1 $ terms in
Eq.~(\ref{eq:STRONG}), employing only the lowest partial-wave
contribution for $G(s,l)$. Using Eq.~(\ref{eq:GGG})  
and the prediction of Ref.~\cite{Capstick:1998uh}, we then can compute
the strong coupling constants for the resonances. The signs of these 
strong coupling constants are determined by  
fitting the experimental data~\cite{Nanova:2008kr,Hleiqawi:2007ad}, 
as will be shown in the next section. We list all the parameters of the 
resonances in Table~\ref{TABLE3}.  
\begin{table}[t]
\begin{tabular}{ccccccccc} \hline\hline
&$\hspace{0.5cm}\mathrm{Resonance}\hspace{0.5cm}$
&$\hspace{0.5cm}G_{s,l}\hspace{0.5cm}$
&$\hspace{0.5cm}g_1\hspace{0.5cm}$
&$\hspace{0.5cm}\Gamma_R\hspace{0.5cm}$
&$\hspace{0.5cm}A_1\hspace{0.5cm}$ 
&$\hspace{0.5cm}A_3\hspace{0.5cm}$ 
&$\hspace{0.5cm}h_1\hspace{0.5cm}$
&$\hspace{0.5cm}h_2\hspace{0.5cm}$\\
\hline
\multirow{1}{*}{$N^*$}
&$D_{13}(2080)$&$-0.5$&$-0.238  $&$300$&$-0.020$&$+0.017$         &$+0.608$&$-0.620$\\
&$S_{11}(2090)$&$-0.9$&$\mp0.909$&$300$&$+0.012$&$\cdot\cdot\cdot$&$+0.055$&$\cdot\cdot\cdot$\\
&$G_{17}(2190)$&$-0.3$&$+5.63   $&$300$&$-0.034$&$+0.028$         &$+7.69 $&$-7.17 $\\
&$D_{15}(2200)$&$+0.2$&$+1.11   $&$300$&$-0.002$&$-0.006$         &$+0.123$&$+0.011$\\
\multirow{1}{*}{$\Delta^*$}
&$S_{31}(2150)$&$-4.8$&$+2.54$  &$300$&$+0.004$&$\cdot\cdot\cdot$&$+0.018$&$\cdot\cdot\cdot$\\
&$G_{37}(2200)$&$+0.5$&$\pm8.32$&$300$&$+0.014$&$-0.004$&$-2.31$ &$+2.47 $\\
&$F_{37}(2390)$&$+0.6$&$+5.02$  &$300$&$+0.024$&$+0.030$&$-1.89$ &$-1.54 $\\
\multirow{1}{*}{$Y^*$}&$\Sigma^*(1385,3/2^+)$
&$\cdot\cdot\cdot$&$\cdot\cdot\cdot$&$\cdot\cdot\cdot$&$\cdot\cdot\cdot$
&$\cdot\cdot\cdot$&$\cdot\cdot\cdot$&$\cdot\cdot\cdot$\\
\hline\hline
\end{tabular}
\caption{Parameters for the resonances in Eqs.~(\ref{eq:LAGEMRE}) and
  (\ref{eq:STRONG}). The decay amplitudes $G(s,l)$ are computed from
  Ref.~\cite{Capstick:1998uh}. The full decay widths $\Gamma_R$ [MeV]
  and helicity amplitudes $A_{1,3}$ [$\mathrm{GeV}^{-\frac{1}{2}}$]
  are taken from the experimental data~\cite{Nakamura:2010zzi} and
  theoretical estimations~\cite{Capstick:1992uc}. The $(+,-)$ sign of
  $g_1$ for $S_{11}$ corresponds to its decay to
  $(K^{*0}\Sigma^+,K^{*+}\Sigma^0)$.}  
\label{TABLE3}
\end{table}

The form factors are included in a gauge-invariant
manner, so that the invariant amplitudes can be expressed as 
\begin{eqnarray}
\mathcal{M}&=&
[\mathcal{M}_{s(N)}^{\mathrm{elec}}+\mathcal{M}_{u(\Sigma)}]F_\mathrm{com}^2 
+ \mathcal{M}_{s(N)}^{\mathrm{mag}} F_N^2
+ \mathcal{M}_{t(K)}F_K^2 + \mathcal{M}_{t(\kappa)} F_\kappa^2 
+ \mathcal{M}_{s(\Delta)} F_\Delta^2
\cr &&
+ \mathcal{M}_{u(\Sigma^*)} F_{\Sigma^*}^2
+ \mathcal{M}_{s(N^*)}F_{N^*}^2
+ \mathcal{M}_{s(\Delta^*)}F_{\Delta^*}^2   
\end{eqnarray}
for the $K^{*0}\Sigma^+$ channel and
\begin{eqnarray}
\mathcal{M}&=&
[\mathcal{M}_{t(K^*)} +
\mathcal{M}_{s(N)}^{\mathrm{elec}} + \mathcal{M}_{\mathrm{c}}]
F_{\mathrm{com}}^2   
+ \mathcal{M}_{s(N)}^{\mathrm{mag}} F_N^2
+ \mathcal{M}_{t(K)}F_K^2 + \mathcal{M}_{t(\kappa)} F_\kappa^2 
+ \mathcal{M}_{s(\Delta)} F_\Delta^2
\cr &&
+ \mathcal{M}_{u(\Lambda)} F_\Lambda^2
+ \mathcal{M}_{u(\Sigma)} F_\Sigma^2
+ \mathcal{M}_{u(\Sigma^*)} F_{\Sigma^*}^2
+ \mathcal{M}_{s(N^*)}F_{N^*}^2
+ \mathcal{M}_{s(\Delta^*)}F_{\Delta^*}^2   
\end{eqnarray}
for the $K^{*+}\Sigma^0$ channel, respectively. The explicit
expressions for each invariant amplitude can be found in Appendix. 
The common form factor $F_{\mathrm{com}}$ and those for the off-mass
shell meson $(\Phi)$ and baryon $(B)$ vertices are written generically
as  
\begin{equation}
\label{eq:FF}
F_{\mathrm{com}}
=F_NF_{\Sigma(K^*)}-F_N-F_{\Sigma(K^*)}, 
\qquad
F_\Phi=\frac{\Lambda^2_{\Phi}-M^2_\Phi}{\Lambda^2_\Phi-q^2},
\qquad
F_B=\frac{\Lambda^4_B}{\Lambda^4_B+(q^2-M^2_B)^2},
\end{equation}
where $q$ denotes the off-shell momentum of the relevant hadron in
each kinematic channel~\cite{Haberzettl:1998eq,
  Davidson:2001rk,Haberzettl:2006bn}. For the mesonic   
$(\Phi=\kappa, K, K^*)$ and baryonic
$(B=N,\Delta,\Lambda,\Sigma,\Sigma^*,R)$ vertices,  
we consider different types of form factors with the cutoff masses 
$\Lambda_\Phi$ and $\Lambda_B$.
\section{Numerical results}
In this section, we present and discuss the numerical results. All the
calculations are performed in the center-of-mass 
(CM) frame. The cutoff masses for the phenomenological form factors
in Eq.~(\ref{eq:FF}) are determined to reproduce the experimental data
for the total and differential cross sections for the $K^{*0}\Sigma^+$
channel from the CBELSA/TAPS~\cite{Nanova:2008kr} 
and CLAS~\cite{Hleiqawi:2007ad} collaborations. 
The determined cutoff masses are listed in Table~\ref{TABLE4}. 
\begin{table}[h]
\begin{tabular}{ccc|cccc|ccc} \hline\hline
\multicolumn{3}{c|}{$\Lambda_\Phi$ for $t$-channel}&
\multicolumn{4}{|c|}{$\Lambda_B$ for $s$-channel}&
\multicolumn{3}{|c}{$\Lambda_B$ for $u$-channel}\\
\hline
$\Lambda_{K^*}$&
$\Lambda_{K}$&
$\Lambda_{\kappa}$&
$\Lambda_{N}$&
$\Lambda_{\Delta}$&
$\Lambda_{N^*}$&
$\Lambda_{\Delta^*}$&
$\Lambda_{\Lambda}$&
$\Lambda_{\Sigma}$&
$\Lambda_{\Sigma^*}$\\
\hline
$0.80$ GeV&
$1.15$ GeV&
$1.15$ GeV&
$1.50$ GeV&
$1.50$ GeV&
$1.00$ GeV&
$1.00$ GeV&
$0.70$ GeV&
$0.95$ GeV&
$0.95$ GeV\\
\hline\hline
\end{tabular}
\caption{Cutoff masses for the form factors in Eq.~(\ref{eq:FF}) for
  each channel.}   
\label{TABLE4}
\end{table}

\begin{figure}[ht]
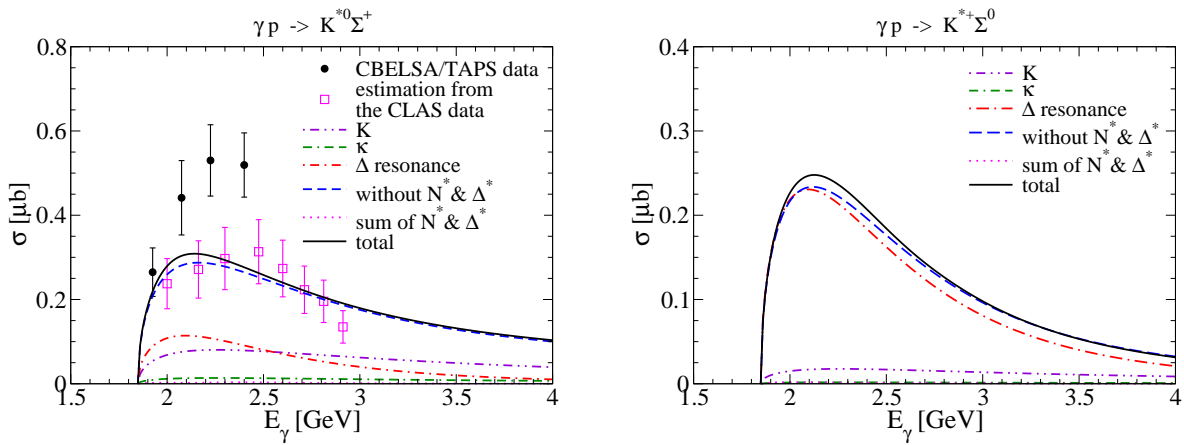

\begin{tabular}{cccc}
\includegraphics[width=7.34cm]{FIG2a.eps} \;\;\;\;\;\;\;\;
\includegraphics[width=7.34cm]{FIG2b.eps}
\end{tabular}
\caption{(Color online) Total cross sections for $\gamma p   \to
  K^{*0}\Sigma^+$ as functions of the photon energy $E_\gamma$ in the 
  left panel. The black circles denote the CBELSA/TAPS
  data~\cite{Nanova:2008kr}, whereas the open squares represent the
  estimated values extracted from the CLAS
  data~\cite{Hleiqawi:2007ad}. The total cross sections for $\gamma p
  \to K^{*+}\Sigma^0$ are given in the right panel with the same
  notation.}         
\label{fig:2}
\end{figure}
We draw the numerical results for the total cross
sections for the $K^{*0}\Sigma^+$ channel in the left panel of
Fig.~\ref{fig:2} in which the $K$-exchange, $\kappa$-exchange, and  
$\Delta$-pole contributions are depicted in dot-dot-dashed, 
dash-dash-dotted, and dot-dashed curves, separately. The solid one
designates the total cross section with all contributions included.  
The black circles denote the CBELSA/TAPS~\cite{Nanova:2008kr} data. We
estimate the total cross sections from the CLAS
data~\cite{Hleiqawi:2007ad} for the differential cross sections, which
are represented by the open squares, based on the interpolating
polynomial method to the fourth order.
Our result shown by the solid line is in a good agreement with 
the CBELSA/TAPS data up to around $E_\gamma\sim2.1$ GeV.
While the present results seem to be underestimated as $E_\gamma$
increases, they are found to be closer to the estimation from the CLAS
data.
We have tried to reproduce the CLAS data rather than those of 
CBELSA/TAPS because there exists more experimental information for 
the wider photon energy region, $E_\gamma$ = (1.925-2.9125).
It turns out that the $K$ exchange and the $\Delta(1232)$-pole
contributions can only describe the experimental data for the $\gamma
p\to K^{*0}\Sigma^+$ total cross section as shown in the dashed curve,
which indicates that the baryon-resonance contributions are almost
negligible. 

In the right panel of Fig.~\ref{fig:2}, we show the results of the
total cross section for the $\gamma p\to K^{*+}\Sigma^0$ process. 
Note that its production strength is a little smaller than
that of the $\gamma p\to K^{*0}\Sigma^+$ one. 
Though the isospin factor of the $K^{*+}\Sigma^0\Delta^+$ vertex
is larger than that of the $K^{*0}\Sigma^+\Delta^+$ one, i.e.
$I_{K^{*+}\Sigma^0\Delta^+} / I_{K^{*0}\Sigma^+\Delta^+}=\sqrt{2}$,
the $t$ channel plays a prominent role in the $K^{*0}\Sigma^+$ process
compared with the $K^{*+}\Sigma^0$ one as shown in Fig.~\ref{fig:2}. 
The other $N^*$, $\Delta^*$ and hyperon resonances have minute effects
on the $K^{*+}\Sigma^0$ production, similar to the
$K^{*0}\Sigma^+$ one. Thus, all other resonances except for
$\Delta(1232)$ seem to be unimportant in describing the {\it
  unpolarized} cross sections for $K^*\Sigma$
photoproduction. However, even though these resonance contributions
are negligibly small, we will see later that they play certain roles in
the polarization observables. In particular, they exhibit more
sensitive angular dependence than other contributions.   
These features are obviously distinguished from the $K^*\Lambda$
photoproduction previously examined in Ref.~\cite{Kim:2011rm}. We 
also verified that with a different set of the strong
coupling constants such as those from the Nijmegen potential
(NSC97f)~\cite{Stoks:1999bz}, we reached the same conclusion.  

\begin{figure}[ht]
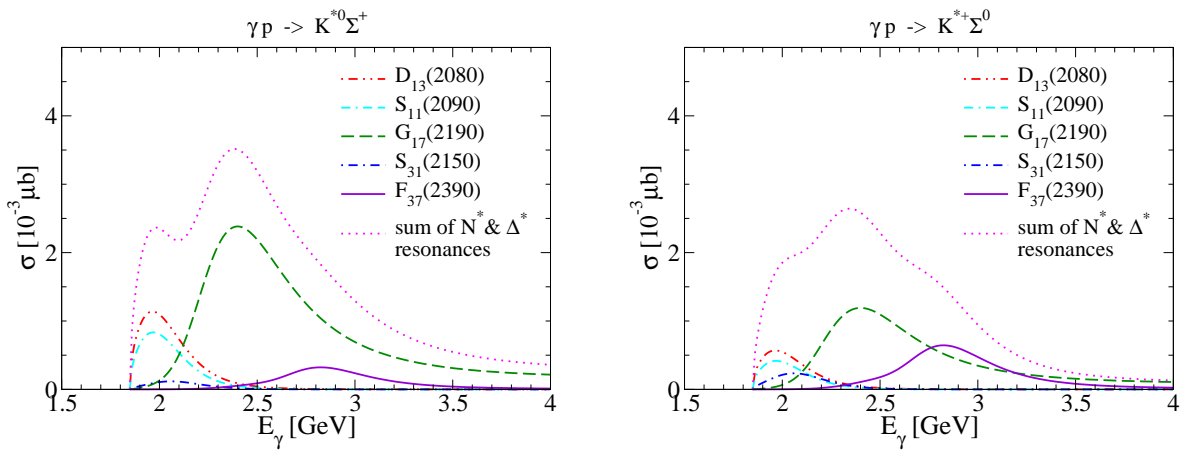

\vspace{0.03cm}
\begin{tabular}{cccc}
\includegraphics[width=7.34cm]{FIG3a.eps} \;\;\;\;\;\;\;\;
\includegraphics[width=7.34cm]{FIG3b.eps}
\end{tabular}
\caption{(Color online) $N^*$ and $\Delta^*$ resonance contributions
  to the total cross sections for $\gamma p \to K^{*0}\Sigma^+$ as
  functions of the photon energy $E_\gamma$ in the left panel and for
  $\gamma p \to K^{*+}\Sigma^0$ in the right panel with the same
  notation, respectively.}          
\label{fig:3}
\end{figure}
Since the $N^*$ and $\Delta^*$ resonances have effects on the
polarization observables as we have mentioned already, it is necessary
to scrutinize them. In Fig.~\ref{fig:3}, we draw each contribution of
the $N^*$ [$D_{13}(2080)$, $S_{11}(2090)$, $G_{17}(2190)$] and $\Delta^*$
[$S_{31}(2150)$, $F_{37}(2390)$] resonances to the total cross section.
Though we computed the contributions of the
$D_{15}(2200)$ and $G_{37}(2200)$, we did not show them
in Fig.~\ref{fig:3}, because they are almost negligible.
As expected, the magnitude of the resonance contributions is about $100$ 
times smaller than that of the Born term contributions. 
This feature of higher $N^*$ and
$\Delta^*$ resonances is very different from the case of $K^*\Lambda$   
photoproduction~\cite{Kim:2011rm}, which ensues from the
fact that the strong coupling constants of $\Sigma$ to these
resonances are much smaller than those of $\Lambda$ to them according
to the SU(6) quark-model calculations~\cite{Capstick:1998uh}. 
Explicitly comparing Table~\ref{TABLE3} in this work with Table~III in
Ref.~\cite{Kim:2011rm}, one can verify, for example,   
$g_{K^*\Sigma D_{13}}/g_{K^*\Lambda D_{13}}\sim1/7$ due to the
different isospin factors.

\begin{figure}[ht]
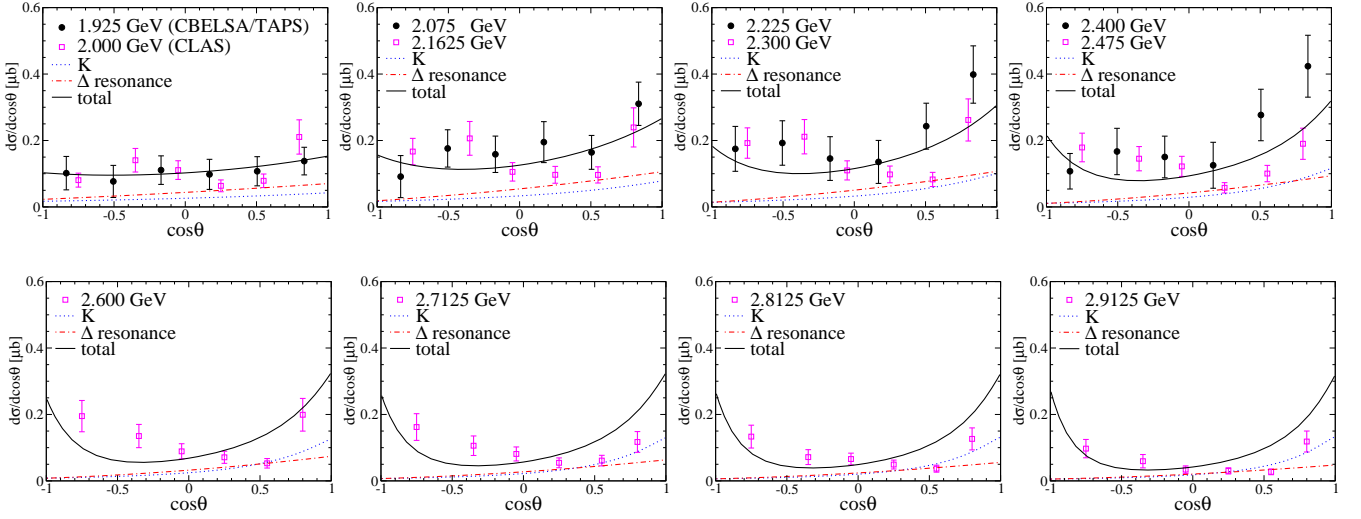

  \centering
\begin{tabular}{cccc}
\includegraphics[width=4.34cm]{FIG4a.eps}
\includegraphics[width=4.34cm]{FIG4b.eps}
\includegraphics[width=4.34cm]{FIG4c.eps}
\includegraphics[width=4.34cm]{FIG4d.eps}
\end{tabular}
\vspace{0.367cm}

\begin{tabular}{cccc}
\includegraphics[width=4.34cm]{FIG4e.eps}
\includegraphics[width=4.34cm]{FIG4f.eps}
\includegraphics[width=4.34cm]{FIG4g.eps}
\includegraphics[width=4.34cm]{FIG4h.eps}
\end{tabular}
\caption{(Color online) Differential cross sections for $\gamma p \to
  K^{*0}\Sigma^+$ as functions of $\cos\theta$ for different photon
  energies ($E_\gamma$) in the range (1.925-2.9125) GeV. The dotted curve
  shows the $t$-channel effects ($K$ and $\kappa$ exchanges), whereas
  the dot-dashed one draws the $\Delta$-pole contribution. The solid  
  one represents the total result. The experimental data of the
  CBELSA/TAPS and CLAS collaborations are taken from
  Refs.~\cite{Nanova:2008kr} and~\cite{Hleiqawi:2007ad},
respectively.}         
\label{fig:4}
\end{figure}
Figure~\ref{fig:4} depicts the numerical results for the
differential cross sections $d\sigma/d\cos\theta$ for the
$K^{*0}\Sigma^+$ channel as functions of $\cos\theta$. The
experimental data are taken from the CBELSA/TAPS~\cite{Nanova:2008kr}
(black circle) and CLAS~\cite{Hleiqawi:2007ad} (open square)
collaborations measured in the range of the photon energy
$E_\gamma$ = (1.925-2.9125) GeV. Note that there is almost no effect
from other $N^*$ and $\Delta^*$ resonances, but our total results
reproduce the data qualitatively well.  
Theoretically, the $t$-channel contributions such
as $\kappa$ and $K$ exchanges enhance the differential cross section
in the forward direction. Although we did not show it explicitly
in the present work, we checked that $K^*$ exchange did not contribute
to the results in the forward direction. We note that the
$\Delta$-pole and $u$-channel Born contributions are responsible for
the enhancement in the backward angle.  

\begin{figure}
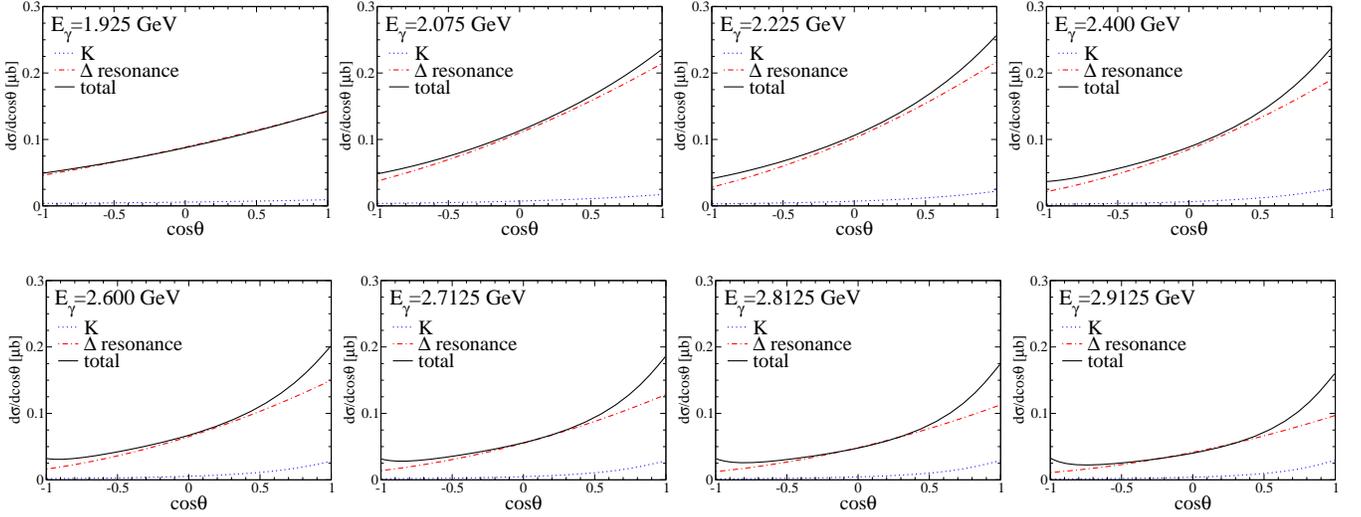

\vspace{0.03cm}
  \centering
\begin{tabular}{cccc}
\includegraphics[width=4.34cm]{FIG5a.eps}
\includegraphics[width=4.34cm]{FIG5b.eps}
\includegraphics[width=4.34cm]{FIG5c.eps}
\includegraphics[width=4.34cm]{FIG5d.eps} 
\vspace{0.367cm}
\end{tabular}
\begin{tabular}{cccc}
\includegraphics[width=4.34cm]{FIG5e.eps}
\includegraphics[width=4.34cm]{FIG5f.eps}
\includegraphics[width=4.34cm]{FIG5g.eps}
\includegraphics[width=4.34cm]{FIG5h.eps}
\end{tabular}
\caption{(Color online) Differential cross sections for $\gamma p \to
  K^{*+}\Sigma^0$ as functions of $\cos\theta$ for different photon
  energies ($E_\gamma$) in the range (1.925-2.9125) GeV. The dotted curve
  shows the $t$-channel effects ($K$ and $\kappa$ exchanges), whereas
  the dot-dashed one draws the $\Delta$-pole contribution. The solid
  one represents the total result.}        
\label{fig:5}
\end{figure}
We also illustrate the differential cross sections for the $\gamma p\to  
K^{*+}\Sigma^0$ process in Fig.~\ref{fig:5} in the same manner as in
Fig.~\ref{fig:4}. As understood from Fig.~\ref{fig:2}, the 
overall strengths of the differential cross sections are smaller than
those of the $K^{*0}\Sigma^+$ channel. Since there are the
$K^*$ exchange, the $\Lambda$ exchange, and the contact term in
addition to other diagrams so as to satisfy the WT identity, the
angular dependence of the differential cross sections for
$K^{*+}\Sigma^0$ photoproduction turns out to be rather different from
those for $K^{*0}\Sigma^+$. Though there is some $t$-channel
contribution to the differential cross section in the forward
direction, the $\Delta$ exchange becomes dominant. 

We are now in a position to discuss the single-polarization
observables. The photon-beam $\Sigma_\gamma$, 
recoil $P_y$, and target $T_y$ asymmetries are defined as
follows~\cite{Titov:2002iv}:  
\begin{equation}
\label{eq:TA}
\Sigma_\gamma\equiv\frac{d\sigma(\epsilon_\perp) -
  d\sigma(\epsilon_\parallel)}{d\sigma_\mathrm{unpol.}},   
\,\,\,\,
P_y\equiv \frac{d\sigma(s^\Sigma_y=\frac{1}{2})
  -d\sigma(s^\Sigma_y=-\frac{1}{2})} 
{d\sigma_\mathrm{unpol.}},
\,\,\,\,
T_y\equiv \frac{d\sigma(s^N_y=\frac{1}{2})-d\sigma(s^N_y=
  -\frac{1}{2})} {d\sigma_\mathrm{unpol.}},
\end{equation}
where $d\sigma_\mathrm{unpol.}$ stands for the unpolarized
differential cross section. These polarization observables
satisfy the following conditions in the collinear limit 
\begin{equation}
\label{eq:CON}
\Sigma_\gamma=P_y=T_y=0\,\,\mathrm{at}\,\,\cos\theta=\pm1.
\end{equation}
Throughout the present work, we define the reaction plane by the
$x$-$z$ axes. Thus, the $y$ axis is perpendicular to the reaction
plane. The photon polarization vectors $\epsilon_{\perp}$ and
$\epsilon_{\parallel}$ are defined in Appendix, while $s^B_y$
indicates the spin of a baryon $B$ along the $y$ direction.  

\begin{figure}[t]
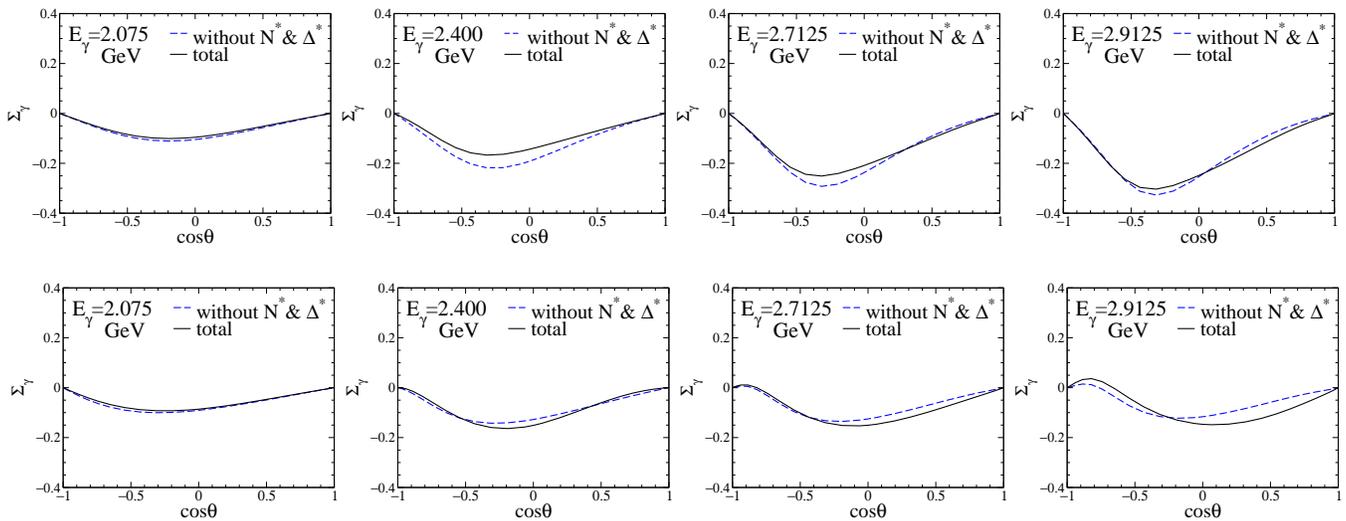

  \centering
\begin{tabular}{cccc}
\includegraphics[width=4.34cm]{FIG6a.eps}
\includegraphics[width=4.34cm]{FIG6b.eps}
\includegraphics[width=4.34cm]{FIG6c.eps}
\includegraphics[width=4.34cm]{FIG6d.eps}
\vspace{0.367cm}
\end{tabular}
\begin{tabular}{cccc}
\includegraphics[width=4.34cm]{FIG6e.eps}
\includegraphics[width=4.34cm]{FIG6f.eps}
\includegraphics[width=4.34cm]{FIG6g.eps}
\includegraphics[width=4.34cm]{FIG6h.eps}
\end{tabular}
\caption{(Color online) In the upper panel, photon-beam asymmetry
  $\Sigma_\gamma$ for $\gamma p \to K^{*0}\Sigma^+$ as functions of
  $\cos\theta$ in the range of $E_\gamma$ = (2.075-2.9125) GeV. The
  solid and dashed curves represent the results with and without the
  resonance   contributions, respectively. In the lower panel, 
  photon-beam asymmetry $\Sigma_\gamma$ for $\gamma p \to
  K^{*+}\Sigma^0$ with the same notation.}        
\label{fig:6}  
\end{figure}
In Fig.~\ref{fig:6}, we depict the numerical results of
$\Sigma_\gamma$ for $K^{*0}\Sigma^+$ in the upper panel and
for $K^{*+}\Sigma^0$ in the lower panel as functions of $\cos\theta$
in the range of $E_\gamma$ = (2.075-2.9125) GeV. It is found that the 
$N^*$ and $\Delta^*$ resonances do not much affect the $\Sigma_\gamma$ 
for both $K^{*+}\Sigma^0$ and $K^{*0}\Sigma^+$ photoproductions, which 
was already seen for the differential cross sections as shown in 
Figs.~\ref{fig:4} and ~\ref{fig:5}. While $\kappa$ and $K$ exchanges 
govern the $K^{*0}\Sigma^+$ production mechanism because of their large 
magnetic couplings, the $\Delta$-pole contribution in the $s$ channel 
pulls down $\Sigma_\gamma$ to the negative direction. The effect of the 
$\Delta$-pole contribution becomes larger as $E_\gamma$ increases. 
The dependence of $\Sigma_\gamma$ on $\cos\theta$ is more complicated 
in the case of the $K^{*+}\Sigma^0$ production, in particular, for higher 
$E_\gamma$, as illustrated in the lower panel of Fig.~\ref{fig:6}. 

\begin{figure}[ht]
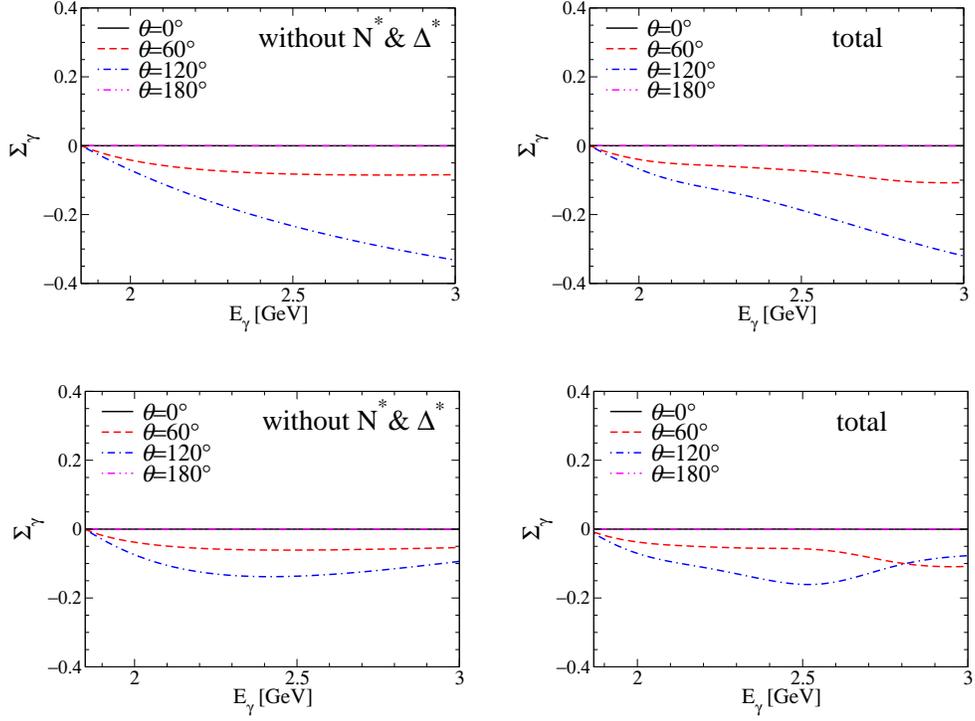

\vspace{0.03cm}
  \centering
\begin{tabular}{cc}
\includegraphics[width=6.cm]{FIG7a.eps} \;\;\;\;\;\;
\includegraphics[width=6.cm]{FIG7b.eps}
\vspace{0.53cm}
\end{tabular} 
\begin{tabular}{cc}
\includegraphics[width=6.cm]{FIG7c.eps} \;\;\;\;\;\;
\includegraphics[width=6.cm]{FIG7d.eps}
\end{tabular}
\caption{(Color online) In the upper panel, photon-beam asymmetries
  $\Sigma_\gamma$ for $\gamma p \to K^{*0}\Sigma^+$ with and
  without $N^*$ and $\Delta^*$ resonances are drawn in order as
  functions of the photon energy $E_\gamma$, the scattering angle
  being changed from $0^\circ$ to 
  $180^\circ$.  In the lower panel, those for $\gamma p \to 
  K^{*+}\Sigma^0$ are shown with and without the resonance
  contributions, respectively.}        
\label{fig:7}  
\end{figure}
In the upper panel of Fig.~\ref{fig:7}, we draw the photon-beam
asymmetries for $K^{*0}\Sigma^+$ photoproduction with
and without $N^*$ and $\Delta^*$ resonances in order as functions of
$E_\gamma$, the scattering angle being varied between 
$\theta=0^\circ$ and $\theta=180^\circ$. In the lower panel, 
$\Sigma_\gamma$ for the $K^{*+}\Sigma^0$ channel is depicted in the
same notation as the $\gamma p \to K^{*0}\Sigma^+$ process. Though 
the effects of the $N^*$ and $\Delta^*$ resonances seem to be small,
one can see a slight change of $\Sigma_\gamma$ as $E_\gamma$
increases. In particular, the influence of the higher resonances is
more clearly revealed in the intermediate angles ($60^\circ \lesssim
\theta \lesssim 120^\circ$), in the case of the $K^{*+}\Sigma^0$ channel. 

\begin{figure}[ht]
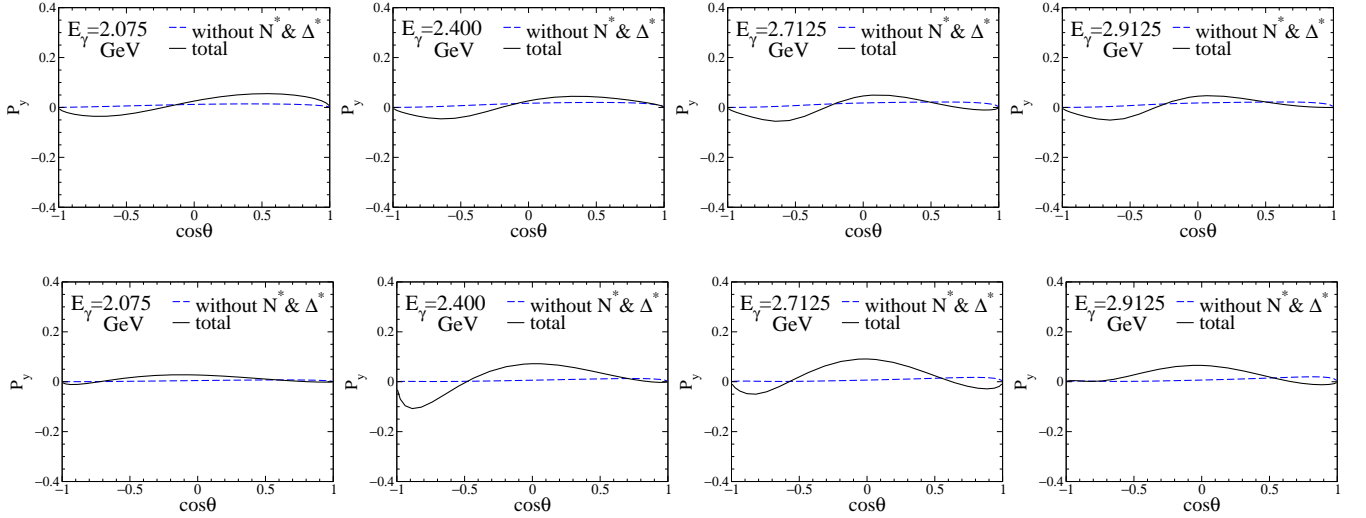

\begin{tabular}{cccc}
\includegraphics[width=4.34cm]{FIG8a.eps}
\includegraphics[width=4.34cm]{FIG8b.eps}
\includegraphics[width=4.34cm]{FIG8c.eps}
\includegraphics[width=4.34cm]{FIG8d.eps}
\vspace{0.367cm}
\end{tabular}
\begin{tabular}{cccc}
\includegraphics[width=4.34cm]{FIG8e.eps}
\includegraphics[width=4.34cm]{FIG8f.eps}
\includegraphics[width=4.34cm]{FIG8g.eps}
\includegraphics[width=4.34cm]{FIG8h.eps}
\end{tabular}
\caption{(Color online) Recoil asymmetries $P_y$ for $K^*\Sigma$
  photoproduction as functions of $\cos\theta$ in the range of the 
  photon energy $E_\gamma$ = 2.075-2.9125 GeV. In the upper and lower
  panels, $P_y$ is drawn for the $K^{*0}\Sigma^+$ and
  $K^{*+}\Sigma^0$ productions, respectively. The solid and dashed
  curves stand for the results with and without the $N^*$ and
  $\Delta^*$ resonances, respectively. }        
\label{fig:8}
\end{figure}
In the upper panel of Fig.~\ref{fig:8}, the recoil asymmetries $P_y$
for $\gamma p \to K^{*0}\Sigma^+$ are presented as functions of
$\cos\theta$ in the range of the photon energy
$E_\gamma$ = (2.075-2.9125) GeV. The solid and dashed curves illustrate
the results of $P_y$ with and without the $N^*$ and $\Delta^*$
resonances. We observe that the higher resonances have some effects on
$P_y$, in contradiction to the case of $\Sigma_\gamma$. 
Since those resonances   
we have considered have rather large spins, their effects on
recoil and target asymmetries defined as the subtraction
between the polarized differential cross sections with opposite spin 
directions of the baryons involved are expected to be
natural. Moreover, the contributions of the $N^*$ resonances are
amplified as $E_\gamma$ increases, as shown in the upper panel of
Fig.~\ref{fig:8}. In the lower panel of Fig.~\ref{fig:8}, $P_y$ for
$K^{*+}\Sigma^0$ photoproduction is depicted. In this case, the
effects of the higher resonances are mild in the lower $E_\gamma$ 
region. However, as $E_\gamma$ increases, $P_y$ starts
to show again some dependence on the scattering angle. 

\begin{figure}[ht]
  \centering
\begin{tabular}{cc}
\includegraphics[width=6.cm]{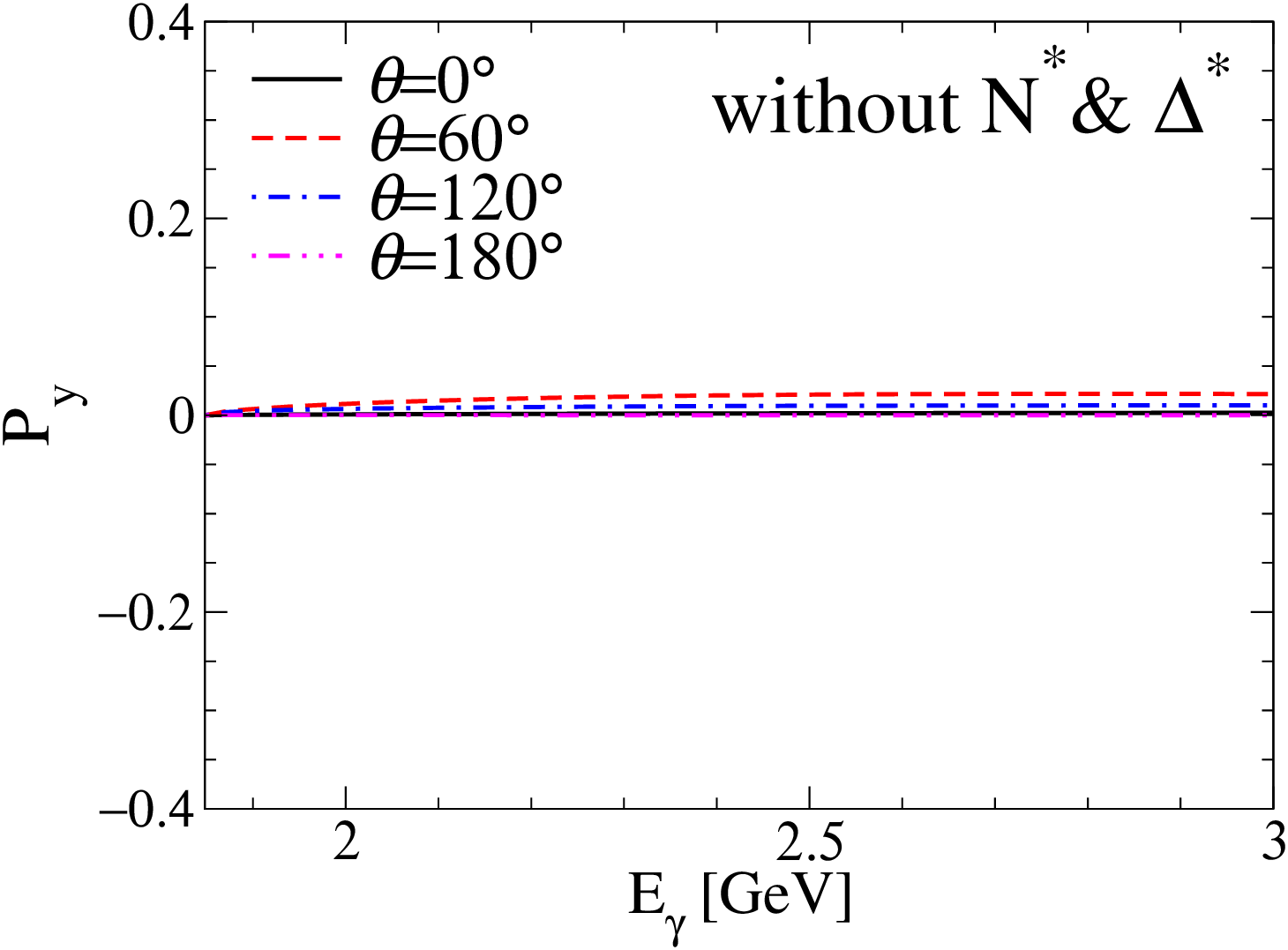} \;\;\;\;\;\;
\includegraphics[width=6.cm]{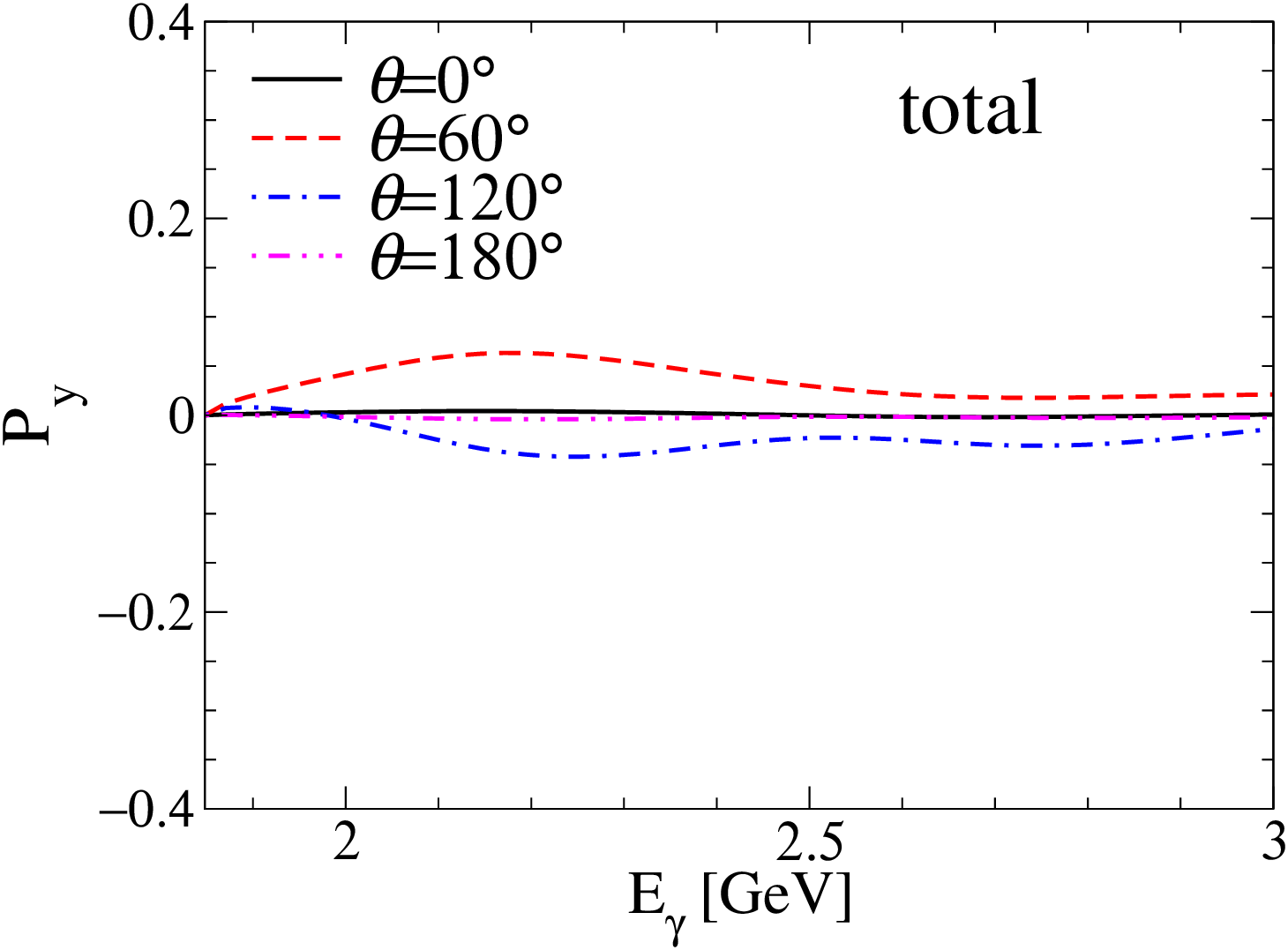}
\vspace{0.53cm}
\end{tabular}
\begin{tabular}{cc}
\includegraphics[width=6.cm]{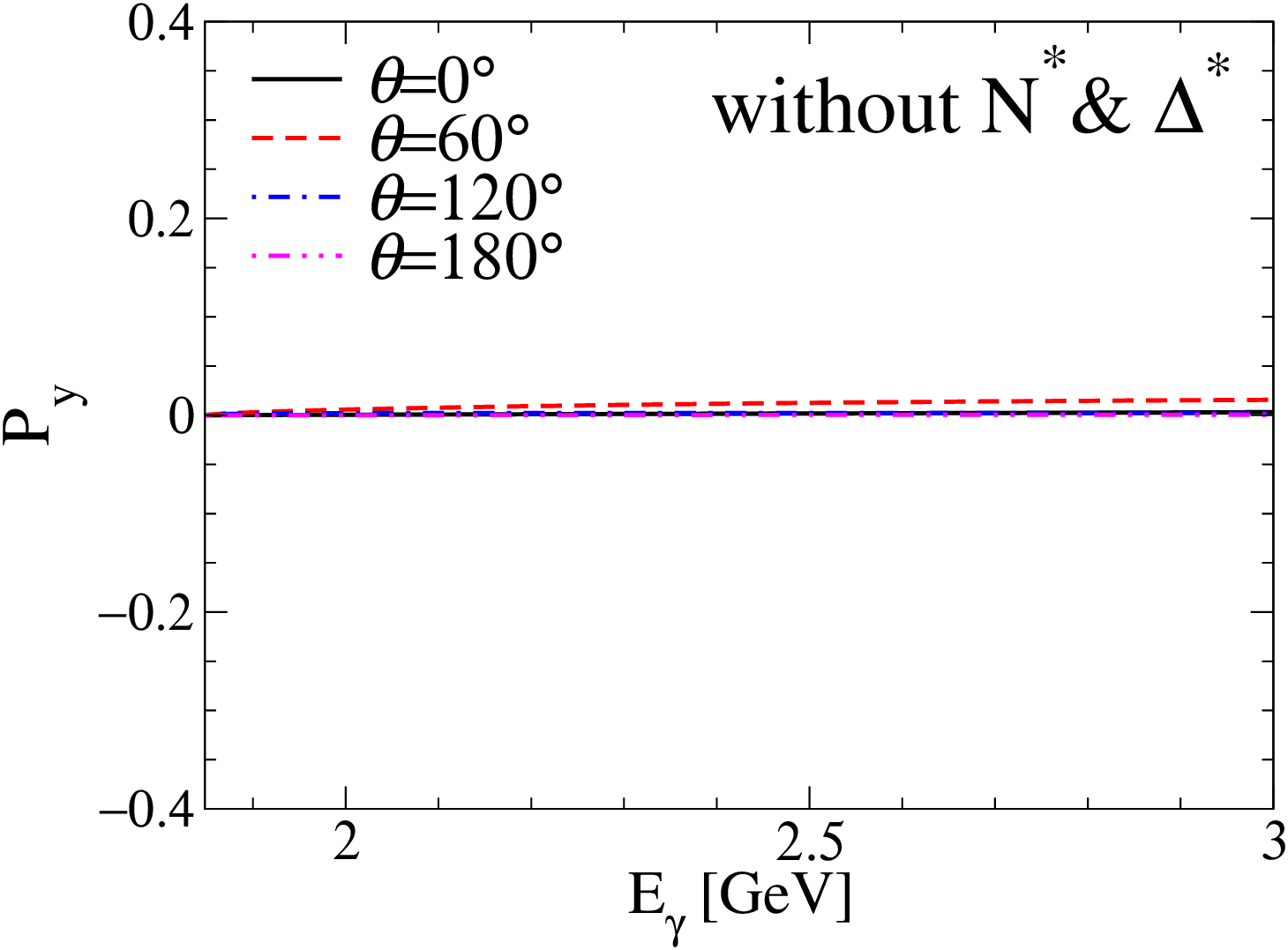} \;\;\;\;\;\;
\includegraphics[width=6.cm]{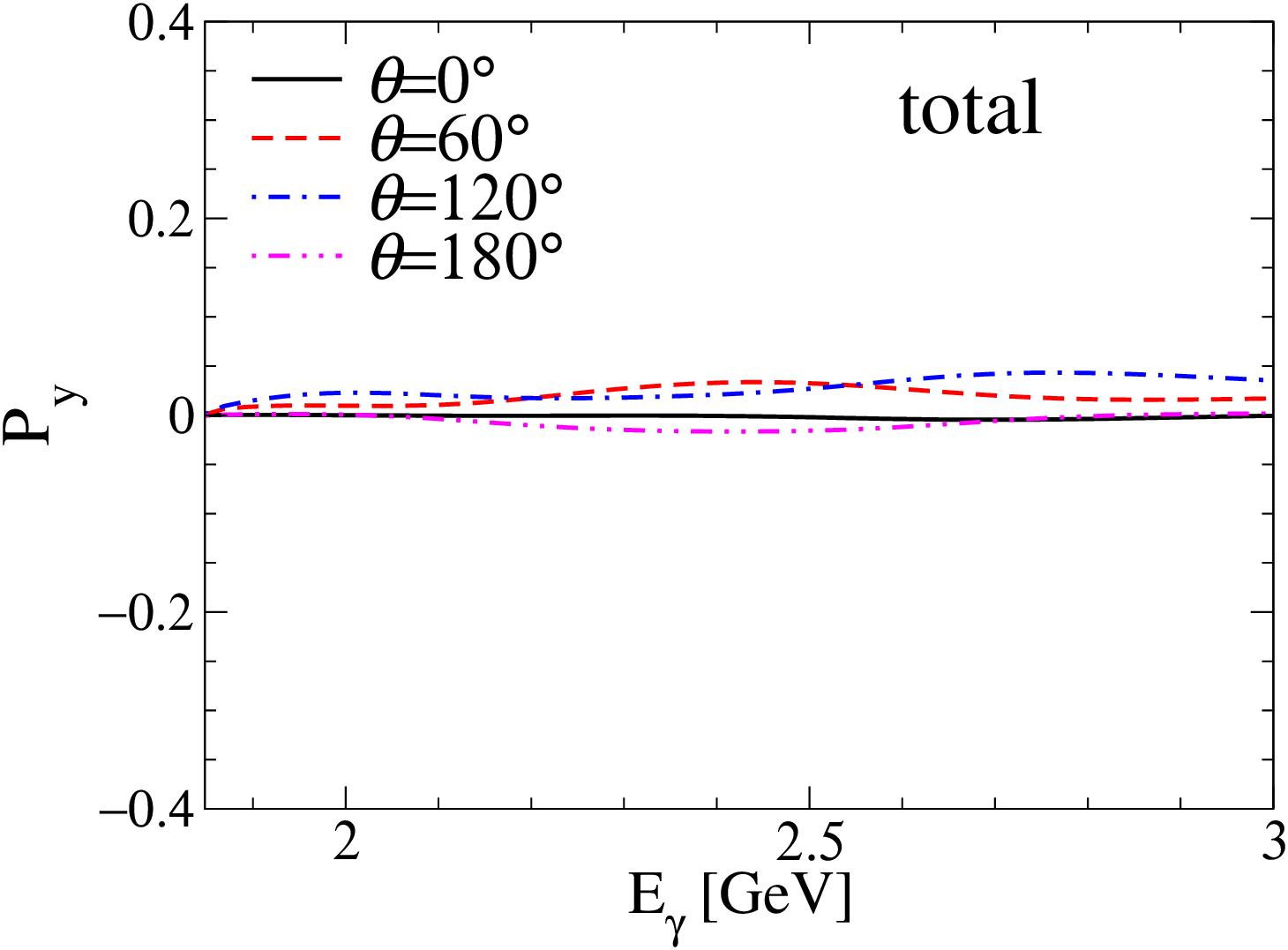}
\end{tabular}
\caption{(Color online) In the upper panel, recoil asymmetries
  $P_y$ for $\gamma p \to K^{*0}\Sigma^+$ with and
  without $N^*$ and $\Delta^*$ resonances are drawn in order as
  functions of the photon energy $E_\gamma$, the scattering angle
  being changed from $0^\circ$ to 
  $180^\circ$.  In the lower panel, those for $\gamma p \to 
  K^{*+}\Sigma^0$ are shown with and without the resonance
  contributions, respectively.}
\label{fig:9}  
\end{figure}
Figure~\ref{fig:9} draws $P_y$ as functions of $E_\gamma$ for the
$K^{*0}\Sigma^+$ channel in the upper panel with and without the $N^*$
and $\Delta^*$ resonances in order and for the $K^{*+}\Sigma^0$
channel in the lower one in the same way. The scattering angle is 
changed from $0^\circ$ to $180^\circ$. When the higher
resonances are turned off, $P_y$ is in general almost independent
of $E_\gamma$. However, including the higher resonances, we
find that $P_y$ at $\theta=60^\circ$ for the $K^{*0}\Sigma^+$ channel 
starts to rise until $E_\gamma\approx 2.2\,\mathrm{GeV}$ and then
falls off slowly, as $E_\gamma$ increases. On the other hand, $P_y$ at
$\theta=60^\circ$ for the $K^{*+}\Sigma^0$ channel begins to increase
around $2.1\,\mathrm{GeV}$ and then saturates around
$2.5\,\mathrm{GeV}$.  

\begin{figure}[ht]
\vspace{0.03cm}
  \centering
\begin{tabular}{cccc}
\includegraphics[width=4.34cm]{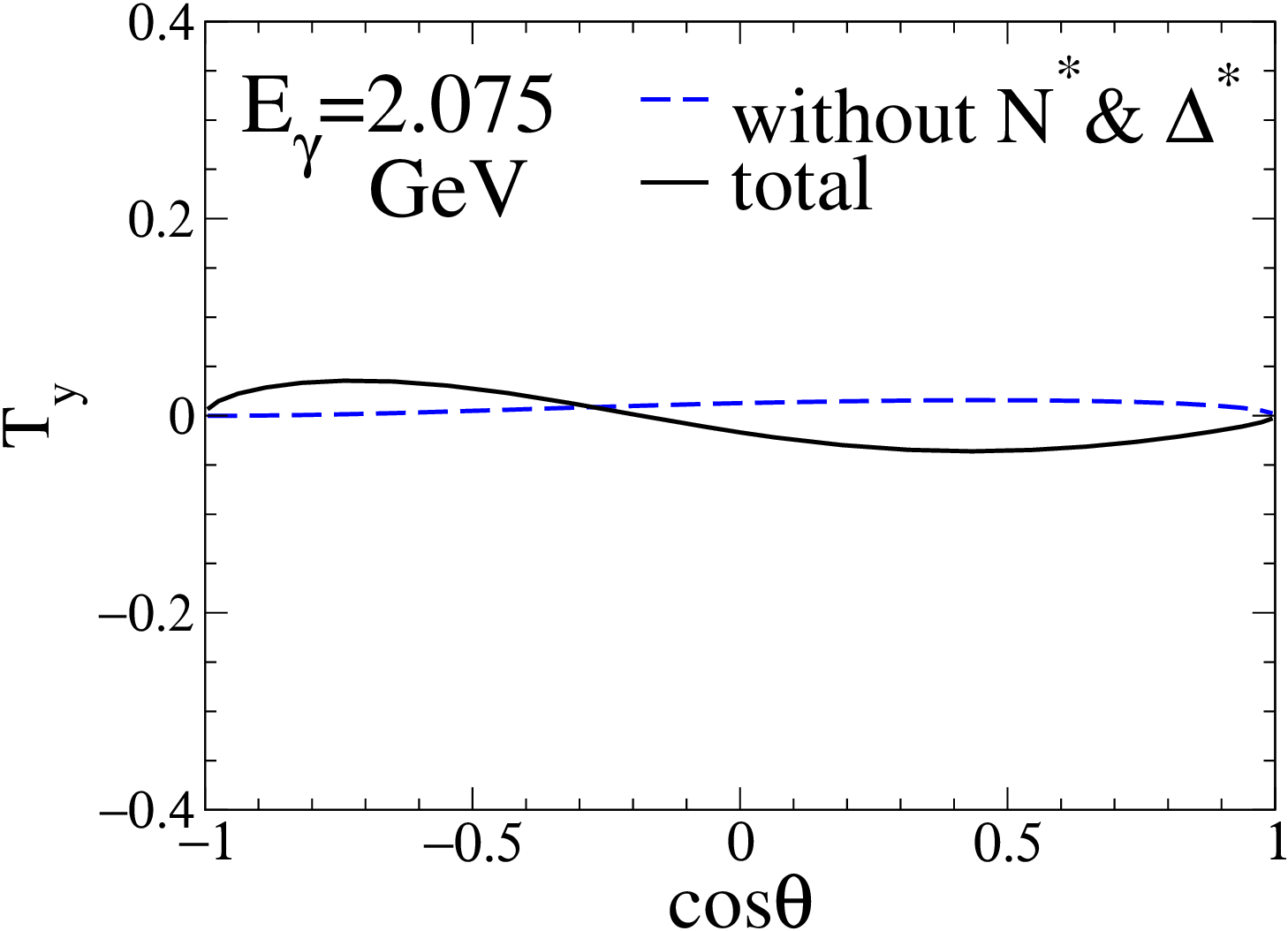}
\includegraphics[width=4.34cm]{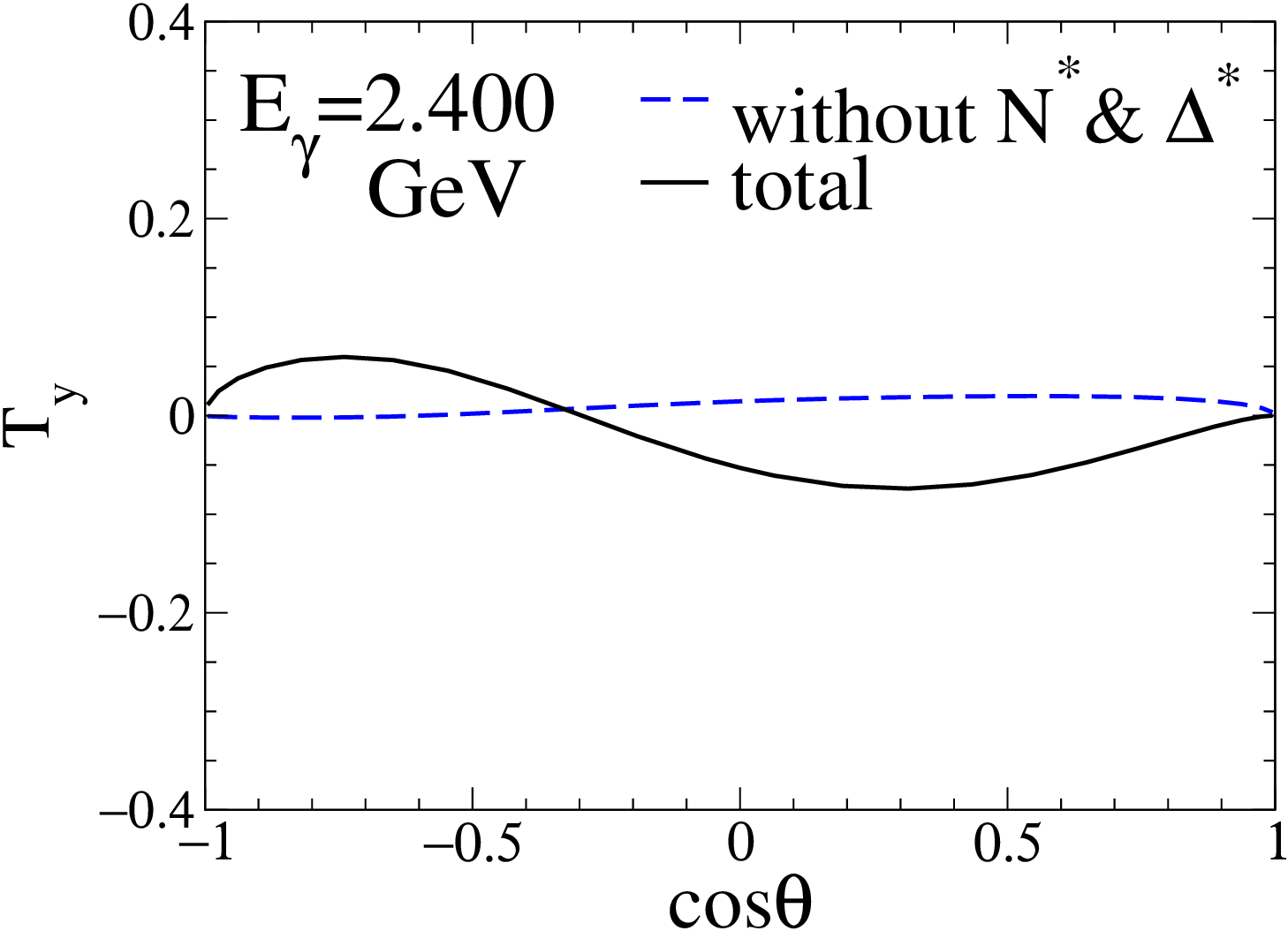}
\includegraphics[width=4.34cm]{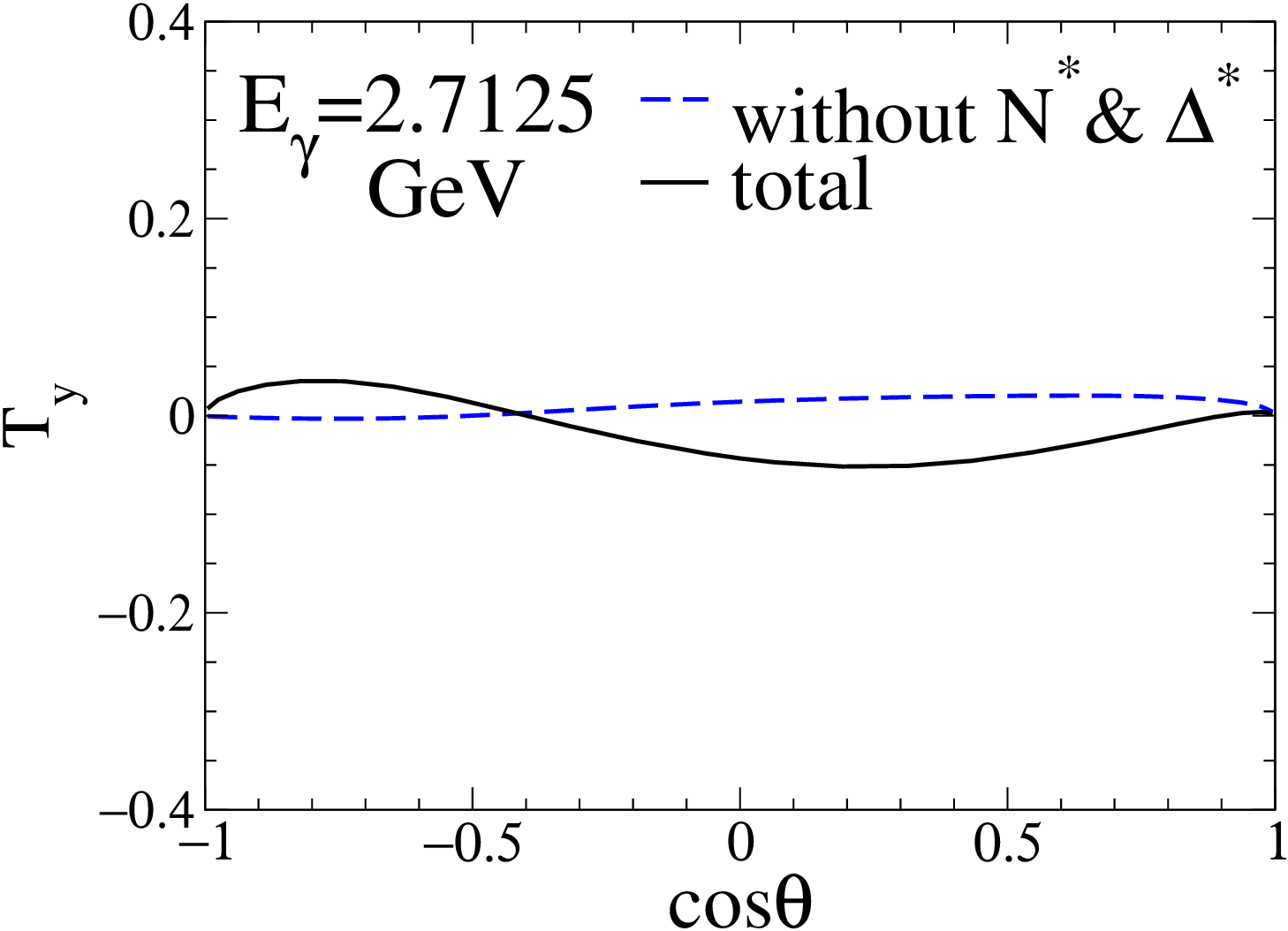}
\includegraphics[width=4.34cm]{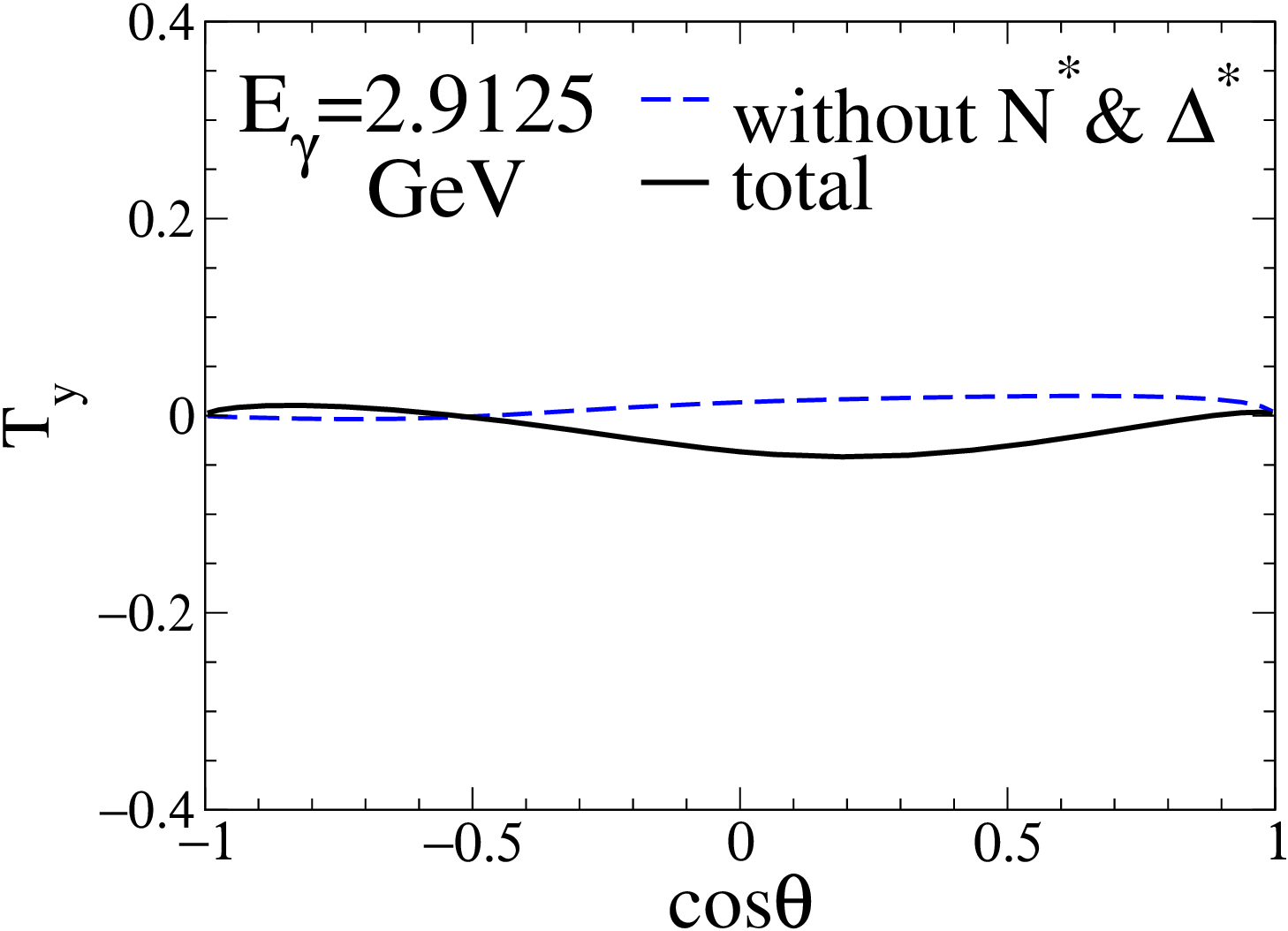}
\vspace{0.367cm}
\end{tabular}
\begin{tabular}{cccc}
\includegraphics[width=4.34cm]{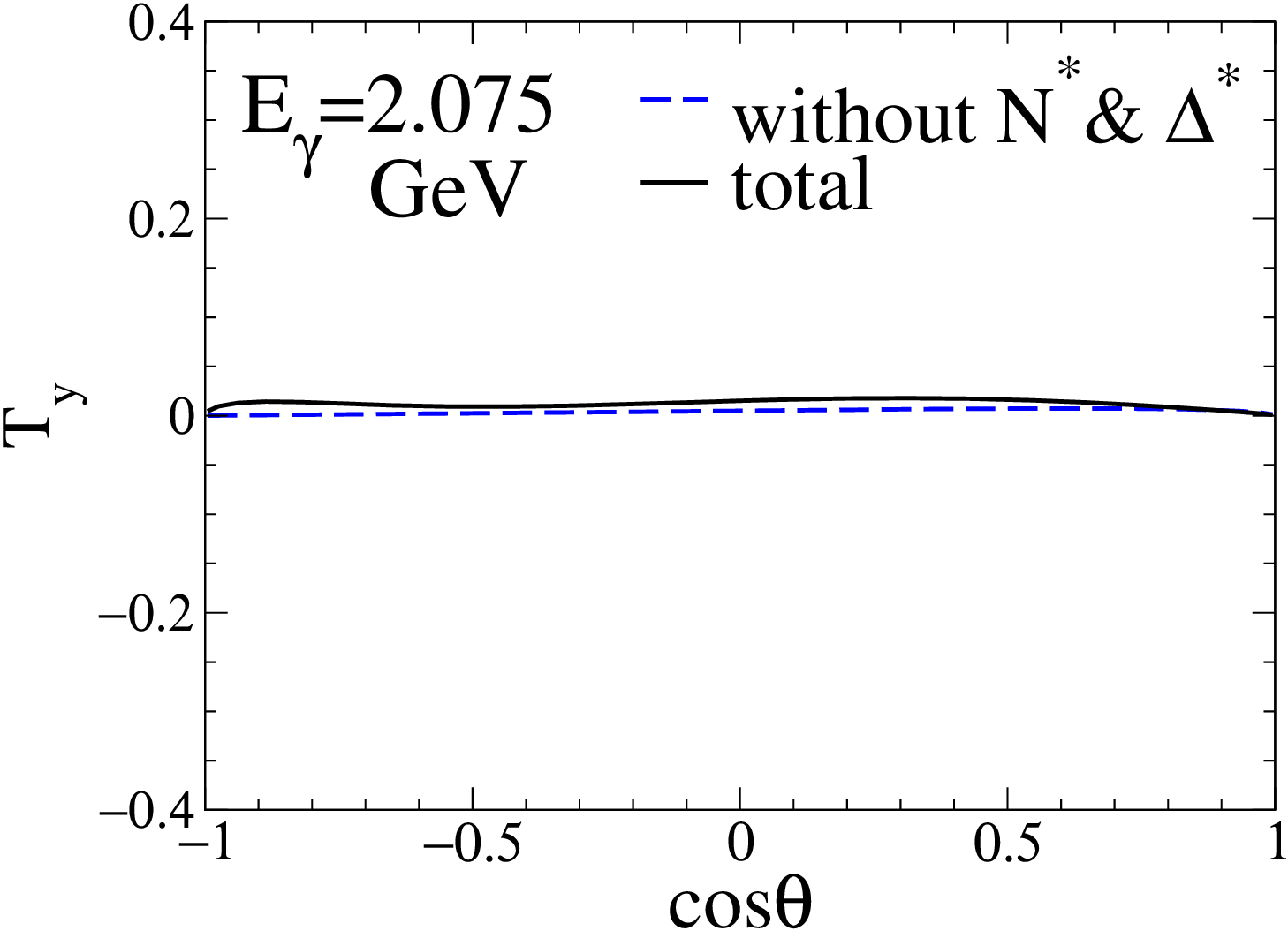}
\includegraphics[width=4.34cm]{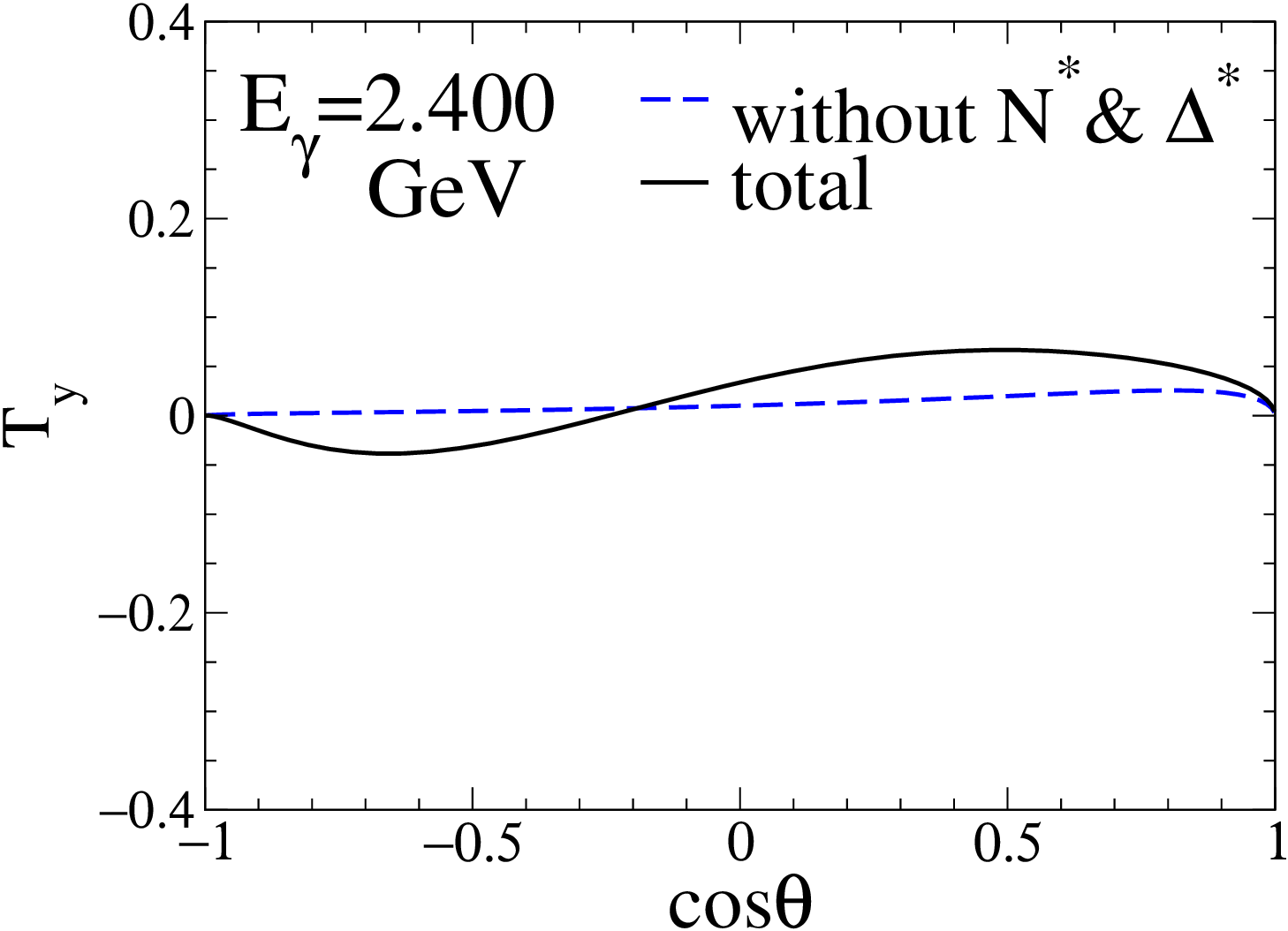}
\includegraphics[width=4.34cm]{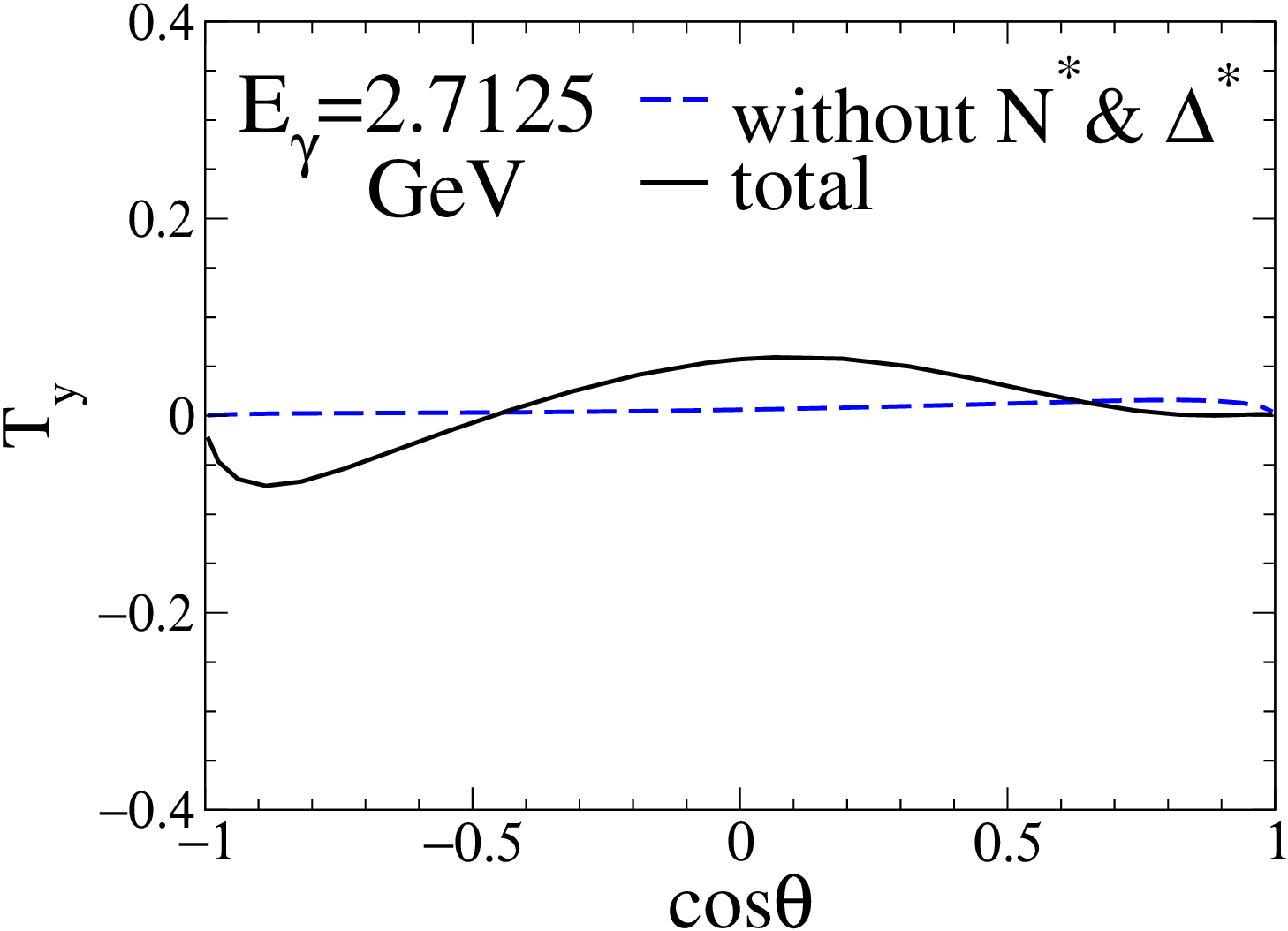}
\includegraphics[width=4.34cm]{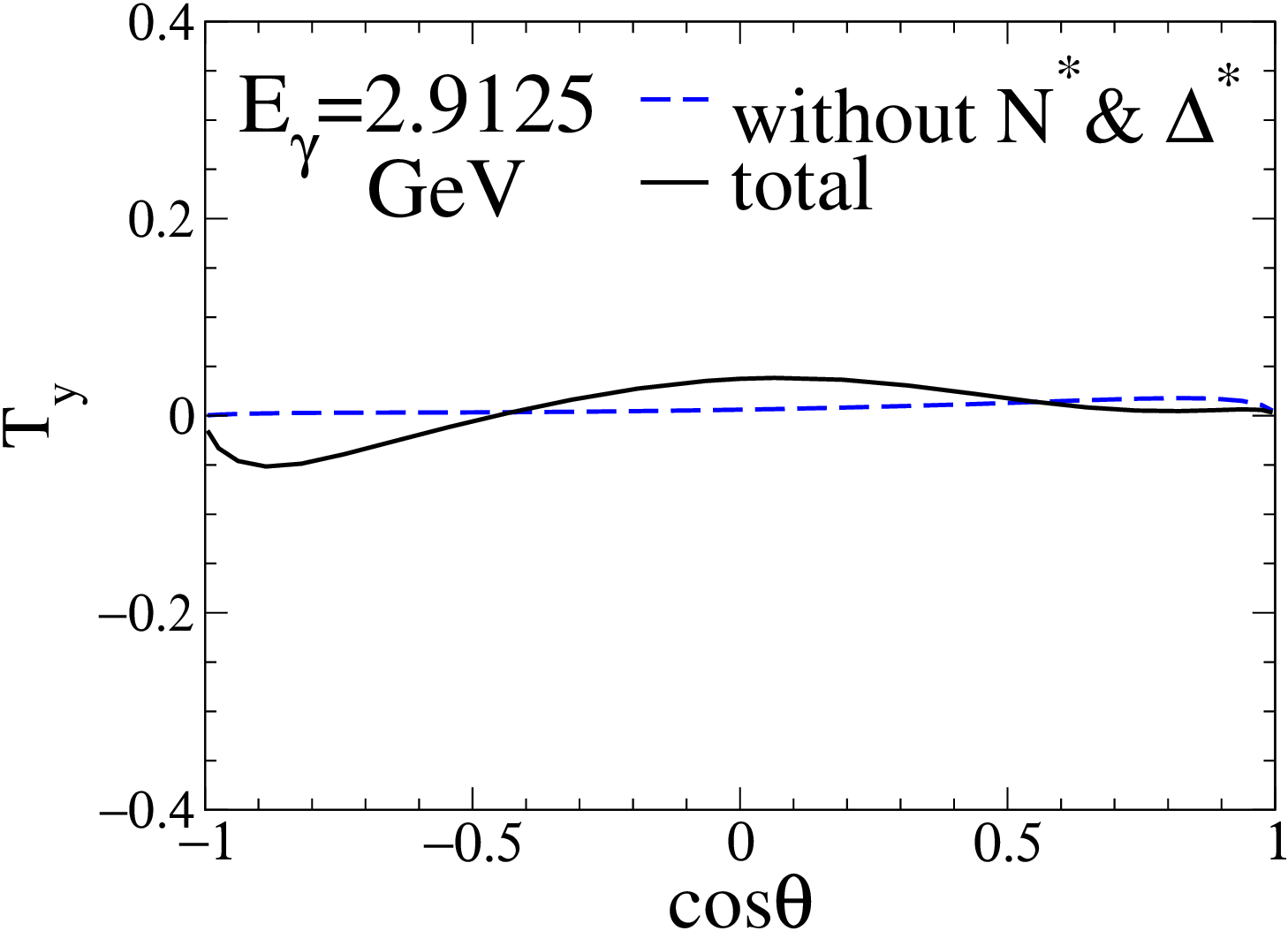}
\end{tabular}
\caption{(Color online) Target asymmetries $T_y$ for $K^*\Sigma$
  photoproduction as functions of $\cos\theta$ in the range of the 
  photon energy $E_\gamma$ = (2.075-2.9125) GeV. In the upper and lower 
  panels, $T_y$ are drawn for the $K^{*0}\Sigma^+$ and
  $K^{*+}\Sigma^0$ productions, respectively. The solid and dashed
  curves stand for the results with and without the $N^*$ and
  $\Delta^*$ resonances, respectively.}
\label{fig:10}  
\end{figure}
\begin{figure}[t]
\begin{tabular}{cc}
\includegraphics[width=6.cm]{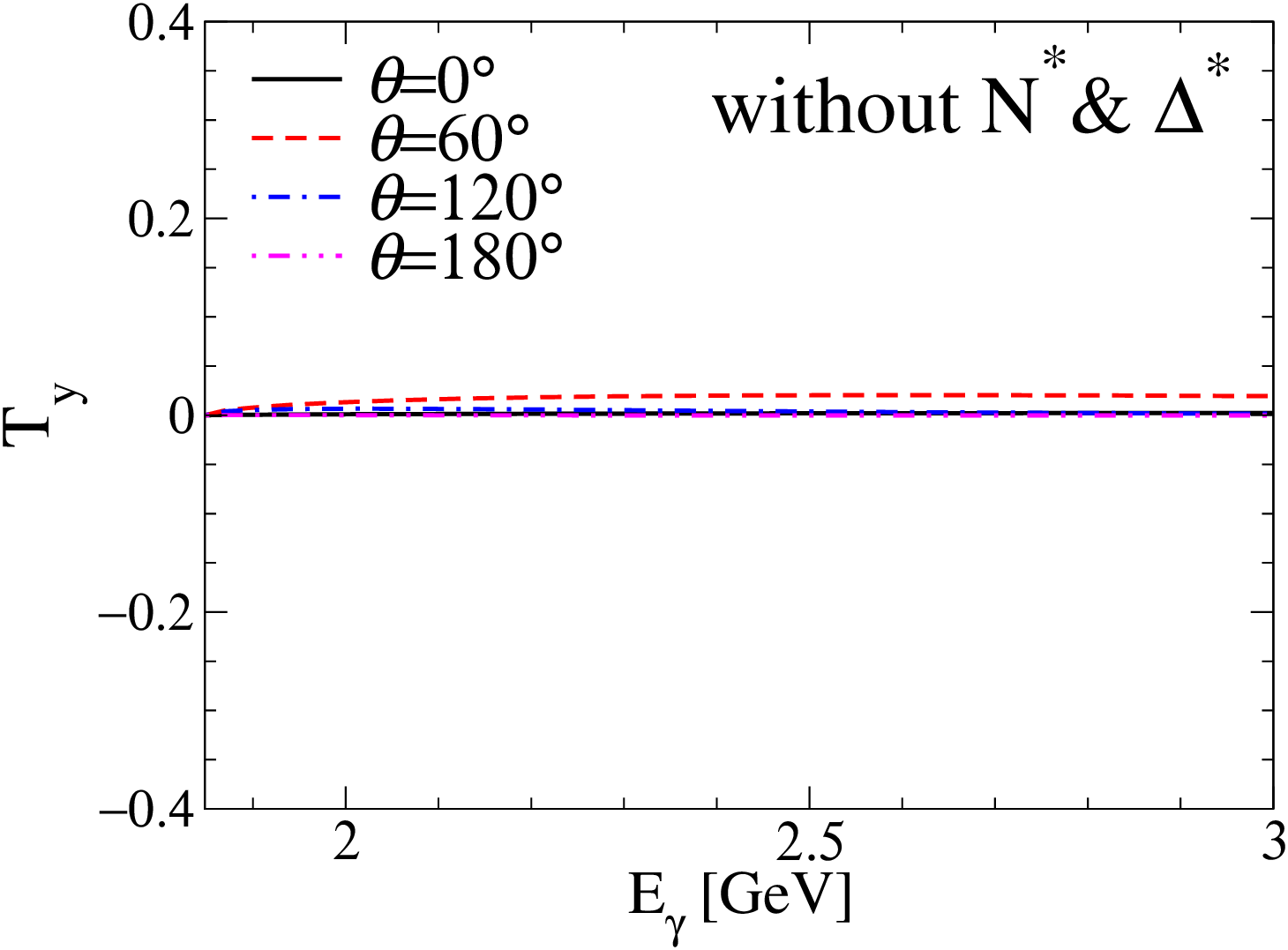}\;\;\;\;\;\;\,\,
\includegraphics[width=6.cm]{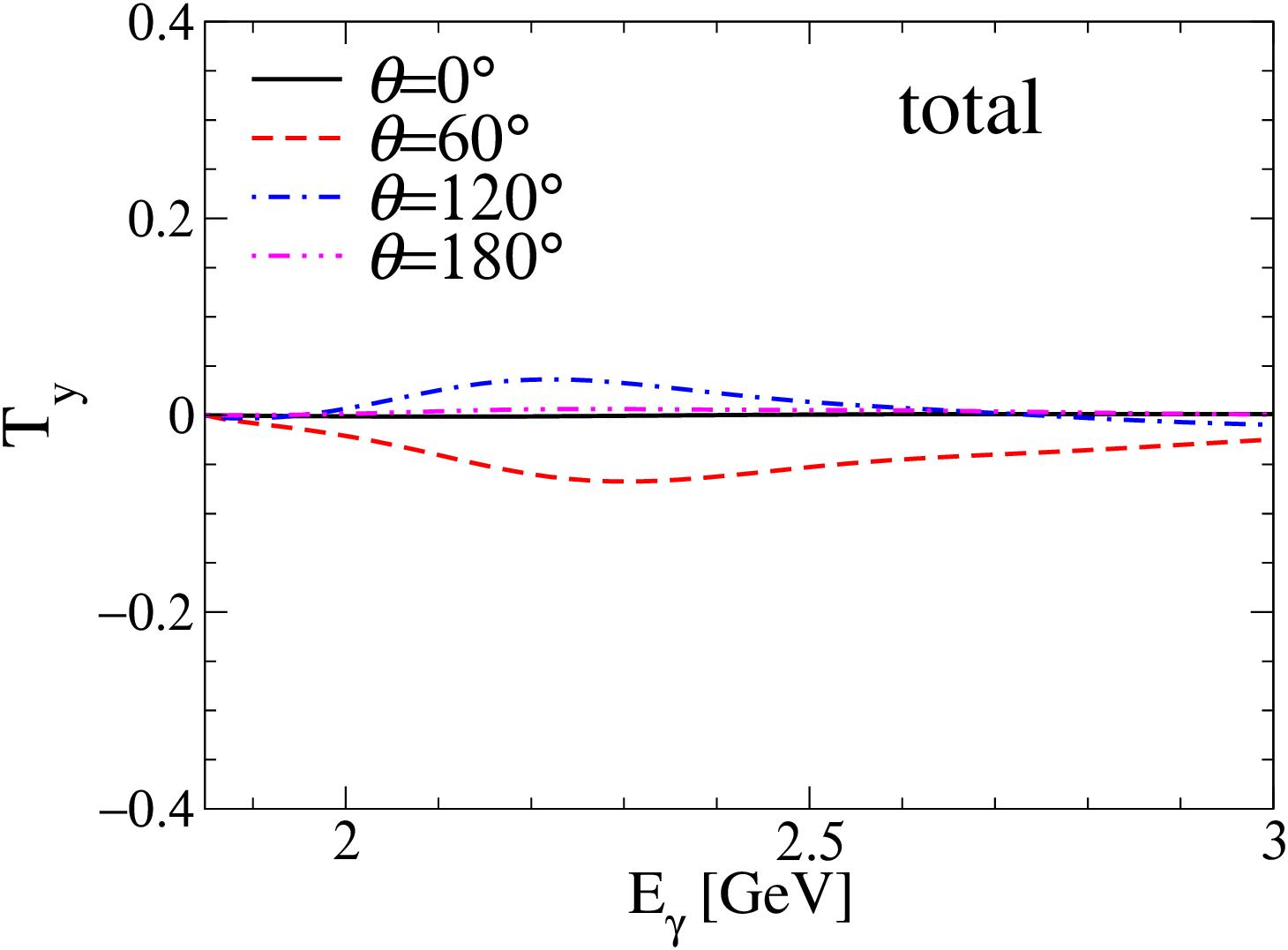}
\vspace{0.53cm}
\end{tabular}
\begin{tabular}{cc}
\includegraphics[width=6.cm]{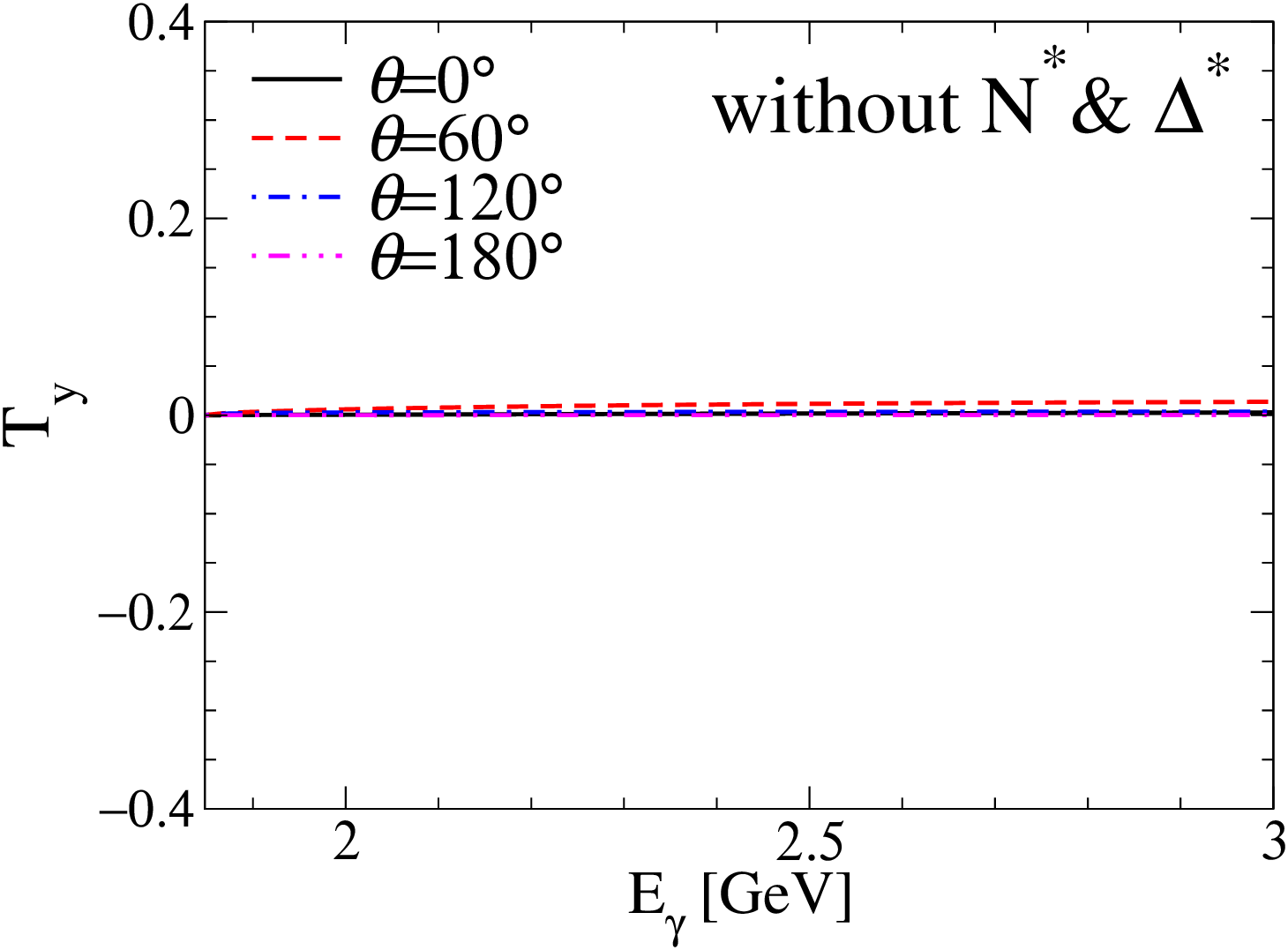}\;\;\;\;\;\;\,\,
\includegraphics[width=6.cm]{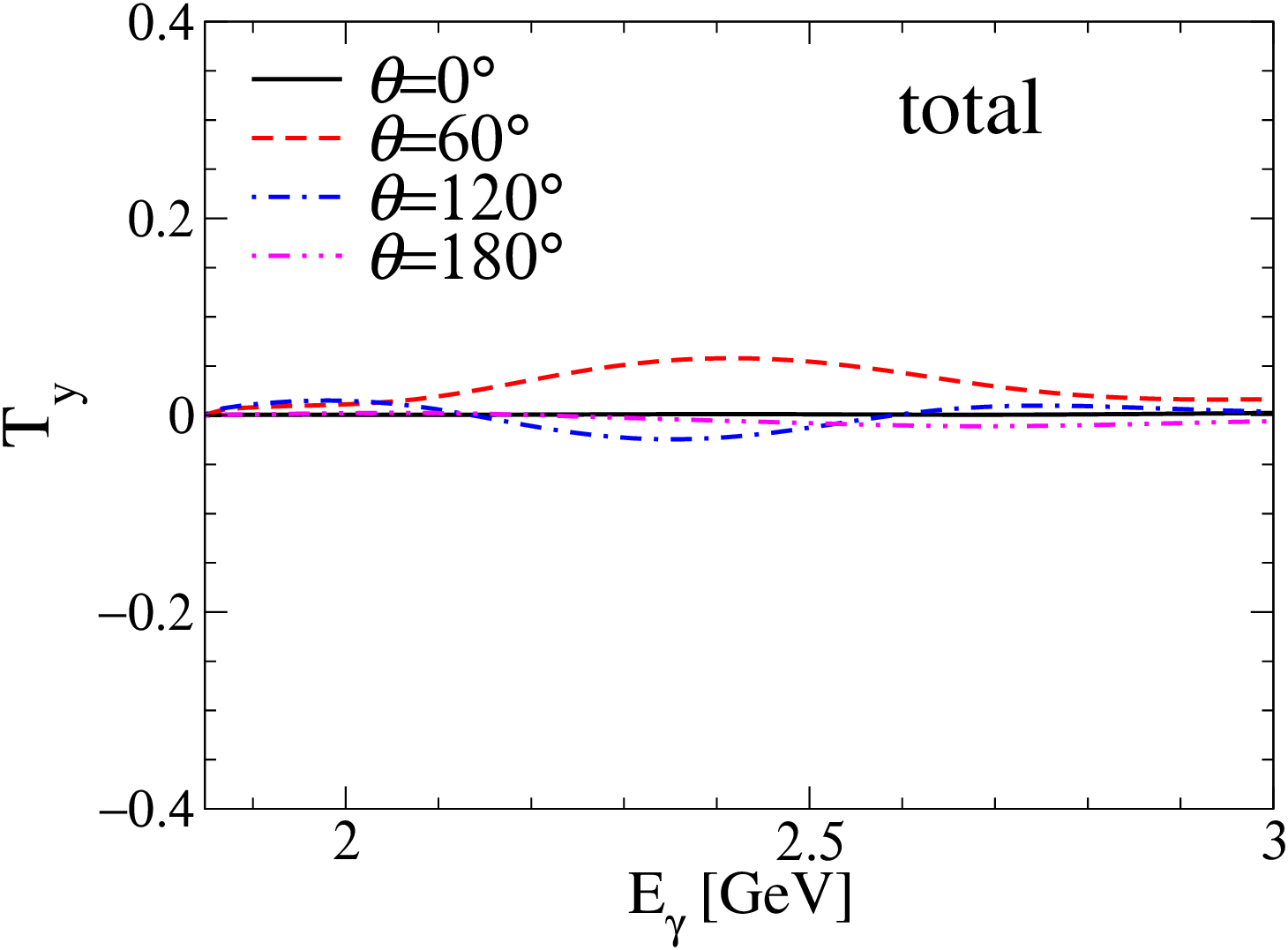}
\end{tabular}
\caption{(Color online) In the upper panel, target asymmetries
  $T_y$ for $\gamma p \to K^{*0}\Sigma^+$ with and
  without $N^*$ and $\Delta^*$ resonances are drawn in order as
  functions of the photon energy $E_\gamma$, the scattering angle
  being changed from $0^\circ$ to 
  $180^\circ$.  In the lower panel, those for $\gamma p \to 
  K^{*+}\Sigma^0$ are shown with and without the resonance
  contributions, respectively.}
\label{fig:11}
\end{figure}

Finally, we provide the numerical results for the target asymmetries
$T_y$ in Fig.~\ref{fig:10} as functions of $\cos\theta$ in the same
manner as in Fig.~\ref{fig:8}. As shown in Fig.~\ref{fig:10}, the
effects of the higher resonances on $T_y$ tend to be very similar to
those on $P_y$. Interestingly, however, we find that the phases of 
the $T_y$ curves for the $K^{*0}\Sigma^+$ and $K^{*+}\Sigma^0$ are
opposite to each other. The dependence of $T_y$ on $E_\gamma$ is shown in
Fig.~\ref{fig:11} in the same way as Fig.~\ref{fig:9}. Again, it turns
out that the higher resonance contributions become obvious around 
$E_\gamma=(2.0\sim2.5)$ GeV, due to the similar reason for $P_y$.   
 
\section{Summary and conclusion}
We have investigated $K^*\Sigma(1193)$ photoproduction, employing
the effective Lagrangian approach at the tree-level Born 
approximation. In addition to the Born diagrams, which satisfy
the WT identity with the phenomenological form factors, we took into
account the baryon-resonance contributions in the $s$ and $u$ channels. 
All the model parameters were determined by using
experimental and theoretical information, reproducing the
available experimental data for the present reaction process. 
We summarize important observations in
the present work as follows:  
\begin{enumerate}[(i)]
\item The unpolarized production strengths for $K^{*0}\Sigma^+$ and
  $K^{*+}\Sigma^0$ photoproductions are negligibly affected by the
  resonance contributions. In other words, the total 
  production rate is dominated by the Born diagrams such as the
  $\Delta$-pole and $K$ exchanges,
  as far as we rely on 
  presently available experimental and theoretical information for the
  resonances taken into account. This tendency is obviously different
  from those for $K\Lambda(1116)$~\cite{Janssen:2001wk} and
  $K^*\Lambda(1116)$~\cite{Kim:2011rm} photoproductions. 
  The total cross section of the $\gamma p \to K^{*+}\Sigma^0$ process
  turns out to be a little smaller than that of
  $\gamma p \to K^{*0}\Sigma^{+}$, because of the isospin factors  
  and the coupling constants.   
\item The angular dependences of the $K^{*0}\Sigma^+$ channel are
  qualitatively well reproduced in comparison with the CLAS
  ~\cite{Hleiqawi:2007ad} experiment data, showing that
  the main dependence comes from the the $\Delta$-pole and $K$
  exchanges. On the contrary, that of the $K^{*+}\Sigma^0$
  channel is dominated by the $\Delta$-pole contribution in the
  $s$ channel, showing rather flat curves. 
 
\item The single-polarization observables such as recoil and target
  asymmetries $P_y$ and $T_y$ are mainly described by the 
  $N^*$ and $\Delta^*$ resonances, though their effects are almost
  invisible in the cross sections. The reason lies in the fact that
  the generic Born and $\Delta(1232)$-exchange contributions play a
  minor role in the polarized observables. On the contrary, it is
  difficult to see the resonance contributions in the
  transversely polarized photon-beam asymmetry $\Sigma_\gamma$, since 
  the electric and magnetic coupling strengths for the $\gamma N R$,
  where $R\equiv(N^*,\Delta^*,Y^*)$, are qualitatively similar to each
  other.   

\item In the present work, we used the experimental data from the 
  Particle Data Group (PDG) book of 2010. In the latest version 
  of 2012, however, the prediction for the electromagnetic properties 
  of some nucleon resonances, i.e. $D_{13}(2080)$, $S_{11}(2090)$, and
  $D_{15}(2200)$, have been updated. 
  In the PDG 2012, these resonances are nominated as $D_{13}(1875)$, 
  $S_{11}(1895)$, $D_{15}(2060)$, and $D_{13}(2120)$. We have repeated our
  calculation by including the latter two resonances, by observing that 
  the effect of the former two is not important because they locate far
  lower than the threshold. In doing so, we used the new values of the 
  photon helicity amplitude, but the same strong coupling constants as 
  those of $D_{15}(2200)$ and $D_{13}(2080)$.
  Then, we come to the conclusion that the effects of resonances are 
  within $15\%$ in comparison with the total results.     
\end{enumerate}

As noted above, $K^*\Sigma(1193)$ photoproduction manifests
obviously different features of the resonance contributions in
comparison with other strangeness productions. The present theoretical   
results, in particular, the single-polarization observables will provide 
useful guides for future experiments in understanding the role of
higher resonances in photoproductions, which can be
measured by the CLAS, LEPS, and CBELSA/TAPS collaborations. The double
polarization observables such as the polarization transport
coefficients $C_{x,y}$~\cite{Bradford:2006ba,Nam:2009cv} are under
progress and appear elsewhere. 

\section*{Acknowledgments}
The authors are grateful to Y.~Oh, K.~Hicks, and W.~Tang for fruitful
discussions and comments for the present work. 
H.Ch.K expresses his gratitude to R. Woloshyn and P. Navratil for their
hospitality during his stay at TRIUMF, where part of the present work
was done. The works of S.H.K. and H.Ch.K. were supported by the Basic
Science Research Program through the National Research Foundation of
Korea funded by the Ministry of Education, Science and Technology
(Grant No. 2012001083). S.H.K. is also supported by the Ministry
of Education, Culture, Science and Technology of Japan. A.H. is
supported in part by the Grant-in-Aid for Scientific Research on
Priority Areas titled ``Elucidation of New Hadrons with a Variety of 
Flavors''(Grant No. E01:21105006). 
\section*{Appendix}
The scattering amplitude for $K^*\Sigma(1193)$ photoproduction can be
written as follows:
\begin{equation}
\label{eq:ScatAmpl}
\mathcal{M} = 
\varepsilon_\nu^* \bar{u}_\Sigma \mathcal{M}^{\mu\nu} u_N \epsilon_\mu,
\end{equation}
where the Dirac spinors of the nucleon and $\Lambda$ are denoted by
$u_N$ and $u_\Sigma$, respectively, and $\epsilon_\mu$ and $\varepsilon_\mu$ 
represent the polarization vectors for the photon and the  
$K^*$, respectively:
\begin{equation}
\label{eq:POLVEC}
\epsilon_\mu=\Big\{
\begin{array}{l}
\epsilon_\parallel=(0,1,0,0)\\
\epsilon_\perp=(0,0,1,0)
\end{array},\,\,\,\,
\varepsilon_\mu=\Bigg\{
\begin{array}{l}
\varepsilon_1=(0,\cos\theta,0,-\sin\theta)\\
\varepsilon_2=(0,0,1,0)\\
\varepsilon_3=\frac{1}{M_{K^*}}(\bm{k}_{K^*},E_{K^*}\sin\theta,
0,E_{K^*}\cos\theta)  
\end{array},
\end{equation}
satisfying $\epsilon^2=\varepsilon^2=-1$, and otherwise zero.

The relevant invariant amplitudes for each kinematic channel without
$(N^*,\Delta^*)$ are given as follows:
\begin{eqnarray}
\label{eq:BORN}
\mathcal{M}_{\mathrm{c}}^{\mu\nu}
&=&-\frac{ie_{K^*} g_{K^*N\Sigma} \kappa_{K^*N\Sigma}}{2M_N}
\sigma^{\mu\nu},
\cr
\mathcal{M}_{t(\kappa)}^{\mu\nu} &=& 
\frac{-2g_{\gamma K^* \kappa} g_{\kappa N\Sigma} }
{t-(M_\kappa-i\Gamma_\kappa/2)^2} 
(k_1 \cdot k_2 g^{\mu\nu} - k_1^\nu k_2^\mu),
\cr
\mathcal{M}_{t(K)}^{\mu\nu} &=&\frac{i g_{\gamma KK^*} g_{KN\Sigma} }{t-M_K^2}  
\epsilon^{\mu\nu\alpha\beta} k_{1\alpha} k_{2\beta} \gamma_5, 
\cr
\mathcal{M}_{t(K^*)}^{\mu\nu} &=& \frac{e_{K^*}  g_{K^* N\Sigma}}
{t-(M_{K^*}-i\Gamma_{K^*}/2)^2}  
\left ( 2k_2^\mu g^{\nu\alpha} - k_2^\alpha g^{\mu\nu} + k_1^\nu
  g^{\mu\alpha} \right ) \left [ g_{\alpha\beta} - \frac{(k_1 -
    k_2)_\alpha 
    (k_1 - k_2)_\beta}{M_{K^*}^2}\right ]
\cr
&\times&\left [ \gamma^\beta -\frac{i\kappa_{K^* N\Sigma}}
{2M_N} \sigma^{\beta \delta}(k_1 - k_2)_\delta \right ], 
\cr
\mathcal{M}_{s(N)}^{\mu\nu} &=&
\frac{g_{K^* N\Sigma} }{s-M_N^2}
\left[ \gamma^\nu - \frac{i\kappa_{K^* N\Sigma} }{2M_N}
  {\sigma^{\nu\alpha}} {k_2}_\alpha  
\right] (\rlap{/}{k}_1 + \rlap{/}{p}_1+M_N)                         
\left[ e_N \gamma^\mu +\frac{ie \kappa_N }{2M_N} \sigma^{\mu\beta}
  {k_1}_\beta  
\right],
\cr
\mathcal{M}_{s(\Delta)}^{\mu\nu} &=& 
\frac{f_{K^*\Delta\Sigma}}{s-(M_\Delta-i\Gamma_\Delta/2)^2} \frac{e}{2M_{K^*}}
\gamma_\rho \gamma_5 (k_2^\beta g^{\nu\rho} - k_2^\rho g^{\nu\beta}) 
\Delta_{\beta\alpha}
\left[ \frac{g_1 }{2M_N} \gamma_\delta - \frac{g_2 }{(2M_N)^2}
  p_{1\delta}  
\right] \gamma_5 (k_1^\alpha g^{\mu\delta} - k_1^\delta g^{\mu\alpha}),
\cr
\mathcal{M}_{u(\Sigma)}^{\mu\nu} &=&
\frac{g_{K^*N\Sigma} }{u-M_\Sigma^2}  
\left[ e_\Sigma \gamma^\mu +\frac{ie \kappa_\Sigma }{2M_N}
  \sigma^{\mu\alpha} {k_1}_\alpha 
\right]  (\rlap{/}{p}_2 - \rlap{/}{k}_1+M_\Sigma)                         
\left[ \gamma^\nu - \frac{i\kappa_{K^* N\Sigma} }{2M_N}
  {\sigma^{\nu\beta}} {k_2}_\beta 
\right].
\end{eqnarray}
Now, we write the corresponding invariant amplitudes for 
$(N^*,\Delta^*)$ for each spin and parity:  
\begin{eqnarray}
\label{eq:RESO}
\mathcal{M}_{s(R)}^{\mu\nu} \left( {\frac{1}{2}^\pm} \right) 
&=& \frac{-ie}{s-M_R^2} \frac{h_{1R_1}}{(2M_N)^2} 
\left[  g_1 \frac{M_{K^*}^2}{M_R \mp M_N} \Gamma^{\nu(\mp)} \mp
       ig_2 \Gamma^{(\mp)} \sigma^{\nu\beta} k_{2\beta} \right]
(\rlap{/}{k_1}+\rlap{/}{p_1}+M_R)
\Gamma^{(\mp)} \sigma^{\mu\alpha} k_{1\alpha},
\cr
\mathcal{M}_{s(R)}^{\mu\nu} \left( {\frac{3}{2}^\pm} \right)  
&=& \frac{e}{s-M_R^2} 
\left[ \frac{g_1 }{2M_N}\Gamma_\rho^{(\pm)} 
      +\frac{g_2 }{(2M_N)^2} p_{2\rho} \Gamma^{(\pm)}  
      -\frac{g_3 }{(2M_N)^2} k_{2\rho} \Gamma^{(\pm)} \right]
( k_2^\beta g^{\nu\rho} - k_2^\rho g^{\nu\beta})    
\cr
&\times& \Delta_{\beta\alpha}(R,k_1 +p_1 ) 
\left[ \frac{\mu_{R_3} }{2M_N} \Gamma_\delta^{(\pm)} \,\mp\,
       \frac{\bar \mu_{R_3} }{(2M_N)^2} \Gamma^{(\pm)} p_{1 \delta} 
     \right](k_1^\alpha g^{\mu \delta} - k_1^\delta g^{\alpha\mu}),   
\cr
\mathcal{M}_{s(R)}^{\mu\nu} \left( {\frac{5}{2}^\pm} \right) 
&=& \frac{e}{s-M_R^2} 
\left[ \frac{g_1 }{(2M_N)^2}\Gamma_\rho^{(\mp)} 
      +\frac{g_2 }{(2M_N)^3} p_{2 \rho} \Gamma^{(\mp)}
      -\frac{g_3 }{(2M_N)^3} k_{2 \rho} \Gamma^{(\mp)} \right] 
k_2^{\beta_2} ( k_2^{\beta_1} g^{\nu\rho} - k_2^\rho
\epsilon^{\nu{\beta_1}})    
\cr
&\times& \Delta_{{\beta_1}{\beta_2};{\alpha_1}{\alpha_2}}(R,k_1 +p_1 )     
\left[ \frac{\mu_{R_5} }{(2M_N)^2} \Gamma^{(\mp)}_\delta \,\pm\,  
       \frac{\bar \mu_{R_5} }{(2M_N)^3} \Gamma^{(\mp)} p_{1 \delta} 
     \right] 
k_1^{\alpha_2} ( k_1^{\alpha_1} g^{\mu\delta} - k_1^\delta
g^{{\alpha_1}\mu} ),  
\cr
\mathcal{M}_{s(R)}^{\mu\nu} \left( {\frac{7}{2}^\pm} \right) &=& 
\frac{e}{s-M_R^2}  
\left[ \frac{g_1 }{(2M_N)^3}\Gamma_\rho^{(\pm)} 
+\frac{g_2 }{(2M_N)^4} p_{2 \rho} \Gamma^{(\pm)}
-\frac{g_3 }{(2M_N)^4} k_{2 \rho} \Gamma^{(\pm)} \right]    
k_2^{\beta_2} k_2^{\beta_3}( k_2^{\beta_1} g^{\nu\rho} 
- k_2^\rho \epsilon^{\nu{\beta_1}})   
\cr
&\times&
\Delta_{{\beta_1}{\beta_2}{\beta_3};{\alpha_1}{\alpha_2}{\alpha_3}} 
(R,k_1 +p_1 )                             
\left[ \frac{\mu_{R_7} }{(2M_N)^3} \Gamma^{(\pm)}_\delta\mp
 \frac{\bar \mu_{R_7} }{(2M_N)^4} \Gamma^{(\pm)} p_{1 \delta} \right] 
k_1^{\alpha_2} k_1^{\alpha_3} ( k_1^{\alpha_1} g^{\mu\delta} -
k_1^\delta g^{{\alpha_1}\mu} ),  
\cr
\mathcal{M}_{u(R)}^{\mu\nu} \left( \frac{3}{2}^+ \right)  &=& 
\frac{f^{(1)}_{K^*N\Sigma^*}}{u-M_{\Sigma^*}^2} \frac{e}{2M_{K^*}} 
\left[ \frac{g_1 }{2M_N} \gamma_\rho + \frac{g_2 }{(2M_N)^2} p_{2\rho}  
\right] (k_1^\beta g^{\rho\mu} - k_1^\rho g^{\beta\mu}) \gamma_5 
\Delta_{\beta\alpha} \gamma_\delta \gamma_5 
(k_2^\alpha g^{\nu\delta} - k_2^\delta g^{\alpha\nu}), 
\cr
&&
\end{eqnarray}
where the definitions for $\Gamma^{(\pm)}$ are given in
Eq.~(\ref{eq:GAMMASPEM}) and each of the decay widths of resonances is
included by replacing $M_R$ in the propagator with
$M_R-i\Gamma_R/2$. The spin-$3/2$, -$5/2$ and -$7/2$ Rarita-Schwinger
spin projections in Eqs.~(\ref{eq:BORN}) and (\ref{eq:RESO}) are given
by  
\begin{eqnarray}
\label{eq:RSSP}
&&\Delta_{\beta\alpha}(R,p)=(\rlap{/}{p}+M_R)
\left[ - g_{\beta\alpha} +\frac{1}{3}\gamma_\beta \gamma_\alpha +
\frac{1}{3M_R} (\gamma_\beta p_\alpha - \gamma_\alpha p_\beta )
 +\frac{2}{3M^2_R}p_\beta p_\alpha \right],
\cr
&&\Delta_{{\beta_1}{\beta_2};{\alpha_1}{\alpha_2}}(R,p)=(\rlap{/}{p}+M_R)     
\cr
&&\times
\left[\frac{1}{2}( \bar g_{{\beta_1}{\alpha_1}} \bar g_{{\beta_2}{\alpha_2}} 
+\bar g_{{\beta_1}{\alpha_2}} \bar g_{{\beta_2}{\alpha_1}})
-\frac{1}{5} \bar g_{{\beta_1}{\beta_2}} \bar g_{{\alpha_1}{\alpha_2}} 
-\frac{1}{10}( \bar \gamma_{\beta_1} \bar \gamma_{\alpha_1} \bar
g_{{\beta_2}{\alpha_2}}  
+\bar \gamma_{\beta_1} \bar \gamma_{\alpha_2} \bar g_{{\beta_2}{\alpha_1}}
+\bar \gamma_{\beta_2} \bar \gamma_{\alpha_1} \bar g_{{\beta_1}{\alpha_2}} 
+\bar \gamma_{\beta_2} \bar \gamma_{\alpha_2} \bar g_{{\beta_1}{\alpha_1}} ) 
\right],
\cr
&&\Delta_{{\beta_1}{\beta_2}{\beta_3};{\alpha_1}{\alpha_2}{\alpha_3}}
(R,p)=(\rlap{/}{p}+M_R)     
\cr
&&\times
\frac{1}{36}\sum_{P(\alpha),P(\beta)}
\left[-\bar g_{{\beta_1}{\alpha_1}} \bar g_{{\beta_2}{\alpha_2}} 
\bar g_{{\beta_3}{\alpha_3}} +\frac{3}{7}\bar g_{{\beta_1}{\alpha_1}}
\bar g_{{\beta_2}{\beta_3}} \bar g_{{\alpha_2}{\alpha_3}}
 +\frac{3}{7} \bar \gamma_{\beta_1} \bar \gamma_{\alpha_1} 
 \bar g_{{\beta_2}{\alpha_2}} \bar g_{{\beta_3}{\alpha_3}}
 -\frac{3}{35} \bar \gamma_{\beta_1} \bar \gamma_{\alpha_1} 
 \bar g_{{\beta_2}{\beta_3}} \bar g_{{\alpha_2}{\alpha_3}} \right].
\end{eqnarray}
Here, we have used the following notations for convenience:
\begin{equation}
\bar{g}_{\alpha\beta} = g_{\alpha\beta} - \frac{p_\alpha p_\beta}{M^2}, 
\,\,\,\,
\bar{\gamma}_\alpha = \gamma_\alpha - \frac{p_\alpha}{M^2}\rlap{/}{p}.
\end{equation}



\begin{thebibliography}{99}
\bibitem{Bradford:2005pt}
  R.~Bradford {\it et al.} (CLAS Collaboration),
  Phys.\ Rev.\ C {\bf 73}, 035202 (2006).
\bibitem{Tsukada:2007jy}
  K.~Tsukada {\it et al.},
  Phys.\ Rev.\ C {\bf 78}, 014001 (2008).
\bibitem{Watanabe:2006bs}
  T.~Watanabe {\it et al.},
  Phys.\ Lett.\ B {\bf 651}, 269 (2007).
\bibitem{Janssen:2001wk}
  S.~Janssen, J.~Ryckebusch, D.~Debruyne, and T.~Van Cauteren,
  Phys.\ Rev.\ C {\bf 65}, 015201 (2001).
\bibitem{Mart:2011ez}
  T.~Mart,
  Phys.\ Rev.\ C {\bf 83}, 048203 (2011).
\bibitem{Yu:2011fv}
  B.~G.~Yu, T.~K.~Choi, and W.~Kim,
  Phys.\ Lett.\ B {\bf 701}, 332 (2011).
\bibitem{DeCruz:2010ry}
  L.~De Cruz, D.~G.~Ireland, P.~Vancraeyveld, and J.~Ryckebusch,
  Phys.\ Lett.\ B {\bf 694}, 33 (2010).
\bibitem{Corthals:2005ce}
  T.~Corthals, J.~Ryckebusch, and T.~Van Cauteren,
  Phys.\ Rev.\ C {\bf 73}, 045207 (2006).
\bibitem{Corthals:2006nz}
  T.~Corthals, D.~G.~Ireland, T.~Van Cauteren, and J.~Ryckebusch,
  Phys.\ Rev.\ C {\bf 75}, 045204 (2007).
\bibitem{Guo:2006kt}
  L.~Guo and D.~P.~Weygand (CLAS Collaboration),
  arXiv:hep-ex/0601010.
\bibitem{Hicks:2010pg}
  K.~Hicks, D.~Keller, and W.~Tang,
  AIP Conf.\ Proc.\ {\bf 1374}, 177 (2011).
\bibitem{Nanova:2008kr} 
  M.~Nanova {\it et al.} (CBELSA/TAPS Collaboration),
  Eur.\ Phys.\ J.\ A {\bf 35}, 333 (2008).
\bibitem{Hleiqawi:2005sz}
  I.~Hleiqawi and K.~Hicks,
  arXiv:nucl-ex/0512039.
\bibitem{Hleiqawi:2007ad}
  I.~Hleiqawi {\it et al.} (CLAS Collaboration),
  Phys.\ Rev.\ C {\bf 75}, 042201 (2007);
  C {\bf 76}, 039905(E) (2007).
\bibitem{Hwang2012} 
  S.~H.~Hwang {\it et al.} (LEPS Collaboration),
  Phys. Rev. Lett. {\bf 108}, 092001 (2012).
\bibitem{Oh:2006hm}
  Y.~Oh and H.~Kim,
  Phys.\ Rev.\ C {\bf 73}, 065202 (2006).
\bibitem{Oh:2006in}
  Y.~Oh and H.~Kim,
  Phys.\ Rev.\ C {\bf 74}, 015208 (2006).
\bibitem{Ozaki:2009wp}
  S.~Ozaki, H.~Nagahiro, and A.~Hosaka,
  Phys.\ Rev.\ C {\bf 81}, 035206 (2010).
\bibitem{Kim:2011rm}
  S.~H.~Kim, S.~i.~Nam, Y.~Oh, and H.-Ch.~Kim,
  Phys.\ Rev.\ D {\bf 84}, 114023 (2011).
\bibitem{Zhao:2001jw}
  Q.~Zhao, J.~S.~Al-Khalili, and C.~Bennhold,
  Phys.\ Rev.\ C {\bf 64}, 052201 (2001).
\bibitem{Nakamura:2010zzi}
  K.~Nakamura (Particle Data Group),
  J.\ Phys.\ G {\bf 37}, 075021 (2010).
\bibitem{Oh:2008abc}
  Y.~Oh, C.~M.~Ko, and K.~Nakayama, 
  Phys.\ Rev.\ C {\bf 77}, 045204 (2008).
\bibitem{Stoks:1999bz}
  V.~G.~J.~Stoks and T.~A.~Rijken,
  Phys.\ Rev.\ C {\bf 59}, 3009 (1999).
\bibitem{Capstick:1992uc} 
  S.~Capstick,
  Phys.\ Rev.\ D {\bf 46}, 2864 (1992).
\bibitem{Capstick:1998uh} 
  S.~Capstick and W.~Roberts,
  Phys.\ Rev.\ D {\bf 58}, 074011 (1998).
\bibitem{Haberzettl:1998eq}
  H.~Haberzettl, C.~Bennhold, T.~Mart, and T.~Feuster,
  Phys.\ Rev.\ C {\bf 58}, R40 (1998).
\bibitem{Davidson:2001rk}
  R.~M.~Davidson and R.~Workman,
  Phys.\ Rev.\ C {\bf 63}, 025210 (2001).
\bibitem{Haberzettl:2006bn}
  H.~Haberzettl, K.~Nakayama, and S.~Krewald,
  Phys.\ Rev.\ C {\bf 74}, 045202 (2006).
\bibitem{Black:2002ek} 
  D.~Black, M.~Harada, and J.~Schechter,
  Phys.\ Rev.\ Lett.\ {\bf 88}, 181603 (2002).
\bibitem{Rarita:1941mf}
  W.~Rarita and J.~Schwinger,
  Phys.\ Rev.\ {\bf 60}, 61 (1941).
\bibitem{Read:1973ye}
  B.~J.~Read,
  Nucl.\ Phys.\ {\bf B52}, 565 (1973).
\bibitem{Machleidt:1987hj} 
  R.~Machleidt, K.~Holinde, and C.~Elster,
  Phys.\ Rep.\ {\bf 149}, 1 (1987).
\bibitem{Molchanov:2004iq} 
  V.~V.~Molchanov {\it et al.} (SELEX Collaboration),
  Phys.\ Lett.\ B {\bf 590}, 161 (2004).
\bibitem{Hackman:1977am} 
  R.~H.~Hackman, N.~G.~Deshpande, D.~A.~Dicus, and V.~L.~Teplitz,
  Phys.\ Rev.\ D {\bf 18}, 2537 (1978).
\bibitem{Darewych:1983}  
  J.~W.~Darewych, M.~Horbatsch, and R.~Koniuk,
  Phys.\ Rev.\ D {\bf 28}, 1125 (1983).
\bibitem{Schat:1995mt} 
  C.~L.~Schat, C.~Gobbi, and N.~N.~Scoccola,
  Phys.\ Lett.\ B {\bf 356}, 1 (1995).
\bibitem{Abada:1995db} 
  A.~Abada, H.~Weigel, and H.~Reinhardt,
  Phys.\ Lett.\ B {\bf 366}, 26 (1996).  
\bibitem{Wagner:1998bu} 
  G.~Wagner, A.~J.~Buchmann, and A.~Faessler,
  Phys.\ Rev.\ C {\bf 58}, 1745 (1998). 
\bibitem{Kim:2005gz} 
  H.-Ch.~Kim, M.~Polyakov, M.~Praszalowicz, G.-S.~Yang, and K.~Goeke,
  Phys.\ Rev.\ D {\bf 71}, 094023 (2005).
\bibitem{Titov:2002iv} 
  A.~I.~Titov and T.~S.~H.~Lee,
  Phys.\ Rev.\ C {\bf 66}, 015204 (2002).
\bibitem{Bradford:2006ba}
  R.~Bradford {\it et al.} (CLAS Collaboration),
  Phys.\ Rev.\ C {\bf 75}, 035205 (2007).
\bibitem{Nam:2009cv}
  S.~i.~Nam,
  Phys.\ Rev.\ C {\bf 81}, 015201 (2010).
\end{thebibliography}
\end{document}